\documentclass[twocolumn]{aa}
\usepackage{epsfig}
\usepackage{graphicx}
\usepackage[latin1]{inputenc}
\def\simlt{\ \raise -2.truept\hbox{\rlap{\hbox{$\sim$}}\raise5.truept   %
\hbox{$<$}\ }}
\def\simgt{\ \raise -2.truept\hbox{\rlap{\hbox{$\sim$}}\raise5.truept   %
\hbox{$>$}\ }}                                                          %

\def\be{\begin{equation}}
\def\ee{\end{equation}}
\def\newline{\hfil\break}

\def\la{\mathrel{\hbox{\rlap{\hbox{\lower4pt\hbox{$\sim$}}}\hbox{$<$}}}}
\def\ga{\mathrel{\hbox{\rlap{\hbox{\lower4pt\hbox{$\sim$}}}\hbox{$>$}}}}

\begin{document}
\title{Probing the physics and history of cosmic reionization with the Sunyaev-Zel'dovich Effect}
   \author{S. Colafrancesco\inst{1,2}, P. Marchegiani\inst{1,2} and M.S. Emritte\inst{2}}
   \offprints{S. Colafrancesco}
 \institute{INAF - Osservatorio Astronomico di Roma
              via Frascati 33, I-00040 Monteporzio, Italy.
              Email: sergio.colafrancesco@oa-roma.inaf.it
              \and
              School of Physics, University of the Witwatersrand, Private Bag 3, 2050-Johannesburg, South Africa.
              Email: sergio.colafrancesco@wits.ac.za
             }
\date{Received / Accepted }
\authorrunning {S. Colafrancesco et al.}
\titlerunning {Studying DA and EoR with the SZE-21cm effect}
\abstract
  %{} leave it empty if necessary
   {The evolution of the universe during the Dark Ages (DA) and the Epoch of Reonization (EoR) marks an important transition in the history of the universe but it is not yet fully understood.}
   {We study here an alternative technique to probe the DA and EoR that makes use of the Comptonization of the CMB spectrum modified by physical effects occurring during this epoch related to the emergence of the 21-cm radiation background. Inverse Compton scattering of 21-cm photon background by thermal and non-thermal electrons residing in the atmospheres of cosmic structures like galaxy clusters, radiogalaxy lobes and galaxy halos, produces a specific form of Sunyaev-Zel'dovich effect (SZE) that we refer to as SZE-21cm.}
  {We derive the SZE-21cm in a general relativistic approach which is required to describe the correct spectral features of this astrophysical effect. We calculate the spectral features of the thermal and non-thermal SZE-21cm in galaxy clusters and in radiogalaxy lobes, and their dependence on the history of physical mechanisms occurring during the DA and EoR. We study how the spectral shape of the SZE-21cm can be used to establish the global features in the mean 21-cm spectrum generated during and prior to the EoR,  and how it depends on the properties of the (thermal and non-thermal) plasma in cosmic structures.
 }
  {We find that the thermal and non-thermal SZE-21cm have peculiar spectral shapes that allow to investigate the physics and history of the EoR and DA. Its spectrum depends on the gas temperature (for the thermal SZE-21cm) and on the electrons minimum momentum (for the non-thermal SZE-21cm). 
The global SZE-21cm signal can be detected (in $\sim 1000$ hrs) by SKA1-low in the frequency range $\nu \simgt 75-90$ MHz, for clusters in the temperature range 5 to 20 keV, and the difference between the SZE-21cm and the standard SZE can be detected by SKA1 or SKA2 at frequencies depending on the background model and the cluster temperature.
}
  {We have shown that the detection of the SZE-21cm can provide unique information on the DA and EoR, and on the cosmic structures that produce the scattering; the frequencies at which the SZE-21cm shows its main spectral features will indicate the epoch at which the physical processes related to the cosmological 21-cm signal occurred and shed light on the cosmic history during the DA and EoR by using local, well-known cosmic structures like galaxy clusters and radio galaxies.}

 \maketitle
%----------------------------------------------------

\section{Introduction}

Departures of the Cosmic Microwave Background (CMB) frequency spectrum from a pure blackbody encode information about the thermal history of the early universe before the epoch of recombination when it emerged from the last scattering surface.
The evolution of the universe after this epoch proceeds through the period of the Dark Ages (DA) that ends $\sim 400$ million years later, when the first galaxies formed and start emitting ionizing radiation. 

The transition period at the end of the DA marks the Epoch of Reionization (EoR). During this epoch, radiation from the very first luminous sources (e.g., early stars, galaxies, and quasars) succeeded in ionizing the neutral hydrogen gas that had filled the universe since the recombination event (see, e.g., Barkana and Loeb 2001, Loeb and Barkana 2001, Bromm and Larson 2004, Ciardi and Ferrara 2005, Choudhury and Ferrara 2006, Furlanetto et al. 2006, Morales and Wyithe 2010).
The current constraints suggest that the EoR roughly occurs within the redshift range of $z \approx 6 - 20$. This cosmic period  is not yet completely understood and various astrophysical probes have been suggested to shed light on this epoch for early structure formation (see Zaroubi 2013 for a review).

Information from the DA period is not explicitly contained in the CMB because baryonic matter and radiation have already decoupled, and the bulk of  baryonic matter in the universe during this period is in the form of neutral hydrogen gas in the inter galactic medium (IGM). Rather than target observations at the first galaxies and quasars that are the rare, early products of gravitational collapse, it is then necessary to detect directly the presence of the ubiquitous hydrogen gas.
One of the methods of achieving this detection is to search for signatures of the (highly redshifted) 21-cm hyperfine transition line of neutral hydrogen (see, e.g., Loeb \& Zaldarriaga 2004; Cooray 2004; Bharadwaj \& Ali 2004; Carilli et al. 2004; Furlanetto \& Briggs 2004; Furlanetto et al. 2006; Pritchard \& Loeb 2010, 2012; Liu et al. 2013).
The 21-cm signal from the DA would appear as a faint, diffuse background detectable at frequencies below 200 MHz (for redshifts $z > 6$). Thus, measuring the brightness temperature of the redshifted 21-cm background could yield information about both the global and local properties of the IGM.
Determining the average brightness temperature over a large solid angle as a function of redshift would eliminate any dependence on local density perturbations and constrain the history of the neutral fraction of hydrogen in the IGM. 

It has been noted that there are several problems related to the observation of the 21-cm background. Firstly, this signal is faint, of the order of tens of mK relative to the CMB (see, e.g., Furlanetto et al. 2006), and until now only upper limits have been obtained (see, e.g., Paciga et al. 2013, Dillon et al. 2014, Parsons et al. 2014). The second problem is related to the presence of galactic and extragalactic foregrounds whose amplitude can be also about four order of magnitude larger than this signal (see, e.g., de Oliveira-Costa et al. 2008). These problems make difficult to study this signal with the present-day and new generation of radio interferometers, since they are not sensitive to the mean signal, but only to its inhomogeneity, and thus require a very precise calibration and knowledge of foregrounds to remove their contribution (see, e.g., discussion in Furlanetto et al. 2006).

Various methods have been proposed to overcome these problems. One possibility is to study the 21-cm fluctuations to measure the mean background through their redshift-space anisotropies (Barkana \& Loeb 2005a); this method 
%was beyond the capabilities of the first generation of interferometers, but 
can be used with the next generation instruments like the Square Kilometer Array (SKA) (see, e.g., McQuinn et al. 2006). 
A second method is to measure the contrast between the 21-cm signal and the bubbles of ionized plasma present during the EoR, and use their contrast to measure the mean amount of neutral gas (see, e.g., Furlanetto et al. 2006 and references therein).

An alternative method that we want to discuss extensively in this paper is to use the SZE-21cm, i.e. the spectral distortion of the CMB spectrum modified by physical effects occurring during the epoch related to the emergence of the 21-cm radiation background, induced by inverse Compton scattering off the intervening electrons in the atmospheres of various cosmic structures, like galaxy clusters, radiogalaxy lobes and galactic halos.

A preliminary attempt to calculate the SZE-21cm has been presented by Cooray (2006). This calculation turns out to be not adequate for a correct description of the SZE-21cm for two reasons:\\
i) the photon background model used for the modification to the CMB caused by mechanisms working during the DA and EoR is unphysical, because it contains a number of artificial discontinuities, under-resolves the main features of interest at $\nu\sim70$ MHz and contains an unphysical reionization history that produces substantial 21-cm signal down to redshifts $z<2$ (i.e., at frequencies $> 300$ MHz);\\
ii) it is performed in the non-relativistic approximation of the Compton scattering process of CMB photons in the hot intra-cluster medium of galaxy clusters thus neglecting any effect induced by the relativistic corrections to this scattering, by multiple scattering effects and by the scattering of additional non-thermal electrons in clusters, as explicitly reported by Cooray (2006).\\ 
Such problems in the Cooray (2006) calculations lead to an incorrect description of the SZE-21cm that has important consequences in using this cosmological probe. In fact, to take full advantage of the   SZE-21cm study, it is necessary to use a full relativistic formalism, its generalization to any order of magnitude in the plasma optical depth $\tau$ and the possibility to include also the combination of various electron populations residing in cosmic structures (see, e.g., Colafrancesco et al. 2003). It is also necessary to use a wider and more physically motivated set of models for the 21-cm background, including also other physical processes that can change this background, such as the effect of Dark Matter heating. Finally, it is worth considering the effect of changing the redshifts at which the different physical processes took place. 
In this paper we perform such a more complete study following the previous lines of investigation.\\
First, to describe the CMB spectrum modified by the 21-cm cosmological background, 
we use the results of the 21cmFAST code (Mesinger, Furlanetto \& Cen 2011) that include realistic physical effects and also additional mechanisms, such as the heating induced by Dark Matter annihilation (e.g., Valdes et al. 2013; Evoli et al. 2014).\\
Secondly, we perform the calculations in the full relativistic formalism for the derivation of the SZE (see, e.g., Colafrancesco et al. 2003 for details), that is suitable to calculate the SZE-21cm in detail, and to derive the precise information about its spectral properties over a wide frequency range and  in a wide set of cosmic structures.  
This general treatment allows, therefore, to increase both the number and the redshift distribution of objects that can be studied with this method, including galaxy clusters with high temperatures (which are the best targets for maximizing the SZE-21cm signal and are more subject to relativistic effects), with radio halos, cool-cores and other complex morphologies, as well as other extragalactic sources with non-thermal electron distributions such as radio galaxies lobes.

The plan of the paper is the following: in Sect 2 we present the general, full relativistic derivation of the SZE-21cm and the models for the frequency distribution of the global 21-cm background we use in the paper. These are new crucial elements of the derivation of the SZE-21cm that have never been provided up to date. In Sect. 3 we discuss the results of our calculations for various scenarios of the radiation background emerging from the DA and EoR, considering various astrophysically motivated scenarios.
We also discuss here, for the first time, the derivation and the possibility to observe both the thermal and the non-thermal SZE-21cm. We discuss our results in the light of the future radio interferometric experiments like the SKA in Sect.4, and we summarize our conclusions in Sect.5.

Throughout the paper, we use a flat, vacuum--dominated cosmological model with $\Omega_m = 0.315$, $\Omega_{\Lambda} = 0.685$ and $H_0 =67.3$ km s$^{-1}$ Mpc$^{-1}$.

\section{Derivation of the SZE-21cm}

\subsection{General derivation of the SZE for a modified CMB spectrum}

The spectral distortion due to the SZE of the CMB is given in the general form by
\begin{equation}
%I_{mod}(x)=\int_{-\infty}^{+\infty} I_{0,mod}(xe^{-s}) P(s) ds
I(x)=\int_{-\infty}^{+\infty} I_{0}(xe^{-s}) P(s) ds
\label{spettro_risultante}
\end{equation}
(see Colafrancesco et al. 2003 for a general derivation of the SZE), where $x=h\nu/(k T_0)$ is the normalized photon frequency, $T_0$ is the CMB temperature, $P(s)$ is the photon redistribution function (yielding the probability of a logarithmic shift  $s=\ln (\nu'/\nu)$ in the photon frequency due to the inverse Compton scattering process) that depends on the electron spectrum producing the CMB Comptonization, and $I_{0}(x)$ is the specific intensity of the incident CMB radiation field. 

The redistribution function $P(s)$, that contains the relativistic corrections required to describe correctly the Compton scattering produced by high temperature or relativistic electrons, is given by the sum of the probability functions to have $n$ scatterings, $P_n(s)$, weighted by the corresponding Poissonian probability:
\begin{equation}
P(s)=\sum_{n=0}^{+\infty} \frac{e^{-\tau} \tau^n}{n!} P_n(s),
\end{equation}
where the optical depth is given by the integral along the line of sight $\ell$ of the electron density
\begin {equation}
\tau=\sigma_T \int n_e d\ell \; ,
\end{equation}
where $n_e$ is the plasma electron density.
Each function $P_n(s)$ is given by the convolution product of $n$ single scattering probability functions $P_1(s)$:
\begin{equation}
P_n(s)=\underbrace{P_1(s) \otimes \ldots \otimes P_1(s)}_{
 \mbox{n times}},
 \label{eq.pns}
\end{equation}
where
\begin{equation}
P_1(s)=\int_0^\infty f_e(p) P_s(s,p) dp ,
\label{eq.p1s}
\end{equation}
and where $f_e(p)$ is the electron momentum distribution function (normalized as to have $\int_0^\infty f_e(p) dp=1$), and $P_s(s,p)$ is the function that gives the probability to have a frequency shift $s$ by an electron with adimensional momentum $p=\beta \gamma$, and is given by the physics of the inverse Compton scattering process (see, e.g., En\ss lin \& Kaiser 2000, Colafrancesco et al. 2003).

The function $P(s)$ that we use in our approach can be calculated at the desired approximation order in the plasma optical depth $\tau$ or via a general relativistic method by using Fourier transform properties (see Colafrancesco et al. 2003 for details), at variance with the case discussed in Cooray (2006) that is only a non-relativistic approximation for values $\tau \ll 1$.

Once the Comptonized spectrum given by eq.(\ref{spettro_risultante}) is calculated, the general form of the SZE is given by the difference:
\begin{equation}
\Delta I(x)=I(x) - I_0(x).
\label{eq.Deltax}
\end{equation}
For the incoming radiation spectrum $I_0(x)$ it is possible, in our general derivation, to use any radiation field. In the original derivation of the SZE the incoming spectrum is given by the standard CMB spectrum
\begin{equation} \label{spettro_inc}
I_{0,st}(x)=2\frac{(k T_0)^3}{(hc)^2} \frac{x^3}{e^x-1} \; ,
\end{equation}
that, inserted in eq.(\ref{spettro_risultante}) and using eq.(\ref{eq.Deltax}), allows to obtain the standard SZE $\Delta I_{st} (x)$.

Our general derivation allows to use the CMB spectrum modified by other physical effects, such as the possible effect of the photon decay (Colafrancesco \& Marchegiani 2014), the effect of non-planckian deviation of the CMB due to the effect of the plasma frequency in an ionized medium (Colafrancesco, Emritte \& Marchegiani 2015), or -- as we study in this paper -- by the modifications of the CMB provided by mechanisms yielding the 21-cm radiation field.

For the case of the CMB spectrum modified by the effects during the DA and EoR,
the expression of the CMB, written as a function of the frequency $\nu$, is given by
\begin{equation}
I_{0,mod}(\nu)=I_{0,st}(\nu)+\delta I(\nu),
\label{cmb_mod_flux}
\end{equation}
where the modification to the CMB spectrum, $\delta I(\nu)$, can be expressed in terms of brigthness temperature change relative to the CMB, defined as:
\begin{equation}
\delta T (\nu)= \frac{c^2}{2k \nu^2} \delta I(\nu).
\label{brigthness.temp}
\end{equation}
In the next Sect. 2.2 we discuss how to obtain the function $\delta I(\nu)$.
Using eqs. (\ref{spettro_risultante}) and (\ref{eq.Deltax}), the SZE-21cm reads:
\begin{equation}
\Delta I_{mod}(\nu)=I_{mod}(\nu) - I_{0,mod}(\nu) \; .
\label{eq.Deltaxmod}
\end{equation}
In the following, we will express the SZE using the brightness temperature change relative to the CMB:
 \begin{equation}
\Delta T (\nu)= \frac{c^2}{2k \nu^2} \Delta I(\nu),
\label{brigthness.sze}
\end{equation}
that is valid for both the standard, $\Delta T_{st}(\nu)$ and the SZE-21cm, $\Delta T_{mod}(\nu)$.

\subsection{The CMB spectrum modified during the DA and EoR}

The CMB radiation spectrum is modified during the DA and EoR by various physical mechanisms: subsequent to recombination, the temperature of neutral gas is coupled to that of the CMB, and no changes in the CMB spectrum can be observed. At redshifts below 200 the gas cools adiabatically, its temperature drops below that of the CMB, and neutral hydrogen resonantly absorbs CMB photons through the spin-flip transition (Field 1959, Scott and Rees 1990, Loeb and Zaldarriaga 2004).
Heating effects of the neutral gas may also occur at high redshifts. As the first Dark Matter (DM) clumps form in the early Universe, the DM WIMP annihilation can in fact produce a substantial heating of the surrounding IGM (Valdes et al. 2013). At much lower redshifts, gas temperature is also expected to heat up again the IGM as luminous sources turn on and their UV and soft X-ray photons re-ionize and heat the gas (Chen and Miralda-Escude 2004).
An additional spectral signature is also expected from the Ly-$\alpha$ radiation field produced by first sources (Barkana and Loeb 2005b) that is coupled to the CMB spectrum through the Wouthuysen-Field effect (Wouthuysen 1952, Field 1959), producing a suppression of the radiation field (see Furlanetto et al. 2006 for details). 

\begin{figure}[ht]
\begin{center}
%\vspace{-10cm}
{
 \epsfig{file=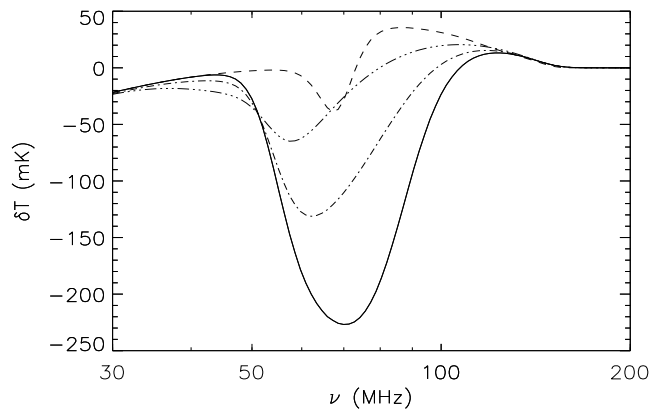,height=6.cm,width=9.cm,angle=0.0}
}
\end{center}
 \caption{\footnotesize{Modified CMB spectrum emerging from the DA and EoR, in units of brightness temperature relative to the CMB (Evoli, private communication): a fiducial model without Dark Matter (solid line; this is our benchmark model), an extreme model without Dark Matter (dashed line),  a fiducial model with Dark Matter with $M_{min}=10^{-3}$ M$_\odot$ (dot-dashed line), where $M_{min}$ is the mass of the smallest DM subhalo, and a fiducial model with Dark Matter with $M_{min}=10^{-6}$ M$_\odot$ (three dots-dashed line).
 }}
 \label{cmb_modified}
\end{figure}

As a result of these physical mechanisms, the CMB spectrum is modified depending on the redshift at which these mechanisms take place.
The spectral shape of the brightness temperature change relative to the CMB (see eq. \ref{brigthness.temp}) is shown in Fig.\ref{cmb_modified}, 
% for the background radiation models used here to calculate the SZE-21cm without and with DM annihilation effects, 
where the background radiation models are calculated with numerical simulations performed using the 21cmFAST code (Evoli, private communication) for different assumptions on the physical processes occurring during the EoR, and without and with DM annihilation effects.\\
The first one (solid line) is a fiducial model without Dark Matter, with standard assumptions on the properties of heating by cosmic structures (see Valdes et al. 2013 and Evoli et al. 2014 for details), without considering the effect of gas collisions which can be observed at frequencies $\nu<30$ MHz, and therefore can not be detected with a ground-based telescope like SKA. This fiducial model takes into account the effects of the Ly-$\alpha$ radiation field at $z\sim30-20$, and 
the effects of UV ionization and X-ray photon heating at $z\sim20-6$.
We use this modified CMB radiation field scheme as a benchmark case for the sake of a general discussion of the SZE-21cm.\\
%
%For the other %more detailed 
%models shown in Fig. \ref{cmb_modified}, we use the results of the 21cm-fast code, with the addition of Dark Matter annihilation heating effects (Valdes et al. 2013), in order to explore more detailed physical effects occurring during the DA and EoR.
%The first model is a fiducial model without Dark Matter, with standard assumptions on the properties of heating by cosmic structures (see Evoli et al. 2014 for details), without considering the effect of gas collisions which can be observed at frequencies $\nu<30$ MHz. 
A second model without Dark Matter that we consider here assumes extreme values for the heating by cosmic structures and, as a result, the deep brightness decrease caused by the coupling of the spin temperature of the IGM with the Ly-$\alpha$ photons is damped, while the emission at higher frequencies is amplified. We finally consider two models with the fiducial parameters and with the heating effects produced by Dark Matter annihilation (Valdes et al. 2013): in these models, the strongest effect is produced by small mass Dark Matter halos, so we consider a model with minimum halo mass $M_{min}=10^{-3}$ M$_\odot$, and one with $M_{min}=10^{-6}$ M$_\odot$, which is more effective in damping the Ly-$\alpha$ coupling effect.
The Dark Matter model used here is a WIMP with mass of 10 GeV and annihilation channel $e^+/e^-$ with cross-section $<\sigma V > = 10^{-26}$ cm$^3$/s.

\subsection{The SZE-21cm spectrum in the benchmark background radiation model}

\begin{figure}[ht]
 \begin{center}
%\vspace{-10cm}
 {
  \epsfig{file=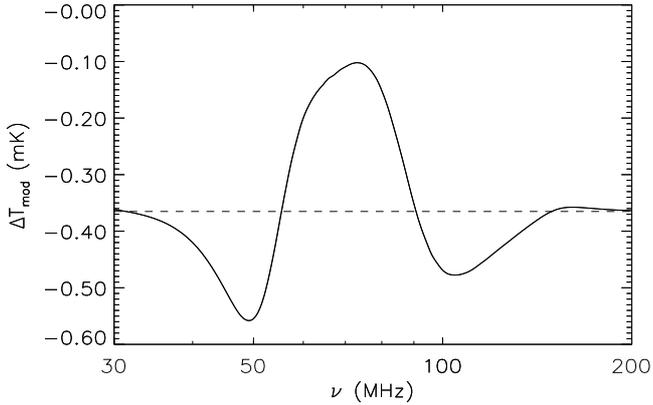,height=6.cm,width=9.cm,angle=0.0}
 }
 \end{center}
  \caption{\footnotesize{The SZE-21cm (in units of brightness temperature relative to the CMB) 
 for a thermal plasma at temperature $kT=7$ keV and with $\tau=5\times10^{-3}$ (solid line). With the dashed line 
 the standard SZE $\Delta T_{st}$ for the same parameters is plotted for comparison.
  }}
  \label{sz_7kev}
 \end{figure}

The modified CMB spectrum (eqs. \ref{cmb_mod_flux} and \ref{brigthness.temp}, where $\delta T$ is shown in Fig.\ref{cmb_modified}) is then scattered by electrons (of both thermal and non-thermal nature) residing in the atmospheres of various cosmic structures, like galaxy clusters, radiogalaxy lobes and galactic halos, and the SZE-21cm is produced. In Figure \ref{sz_7kev} we show an example of the SZE-21cm, $\Delta T_{mod}$, calculated in a galaxy cluster with a temperature of $kT=7$ keV, and using the benchmark model for the modified CMB spectrum shown in Fig.\ref{cmb_modified}.

Figure \ref{sz_7kev} shows that in some frequency bands the SZE-21cm is stronger than the standard one, whereas in other bands it is weaker. This behaviour is mainly related to the curvature of the input spectrum $\delta T$ (see Fig.\ref{cmb_modified}): in the frequency range where the input spectrum has a negative curvature (for $\nu\simlt55$ MHz and $90\simlt\nu\simlt140$ MHz for our fiducial model), the SZE-21cm is smaller than the standard one (i.e. $\Delta T_{mod}-\Delta T_{st}<0$), while at frequencies where the curvature is positive ($55\simlt\nu\simlt90$ MHz and $\nu\simgt140$ MHz) we have $\Delta T_{mod}-\Delta T_{st}>0$. This is due to the fact that the inverse Compton scattering produces a shift in the frequency of photons and, as a consequence, the amplitude of the SZE at a certain frequency depends on the distribution of the photons around that frequency (see, e.g., the shape of the function $P_1(s)$ defined in eq. \ref{eq.p1s} in Colafrancesco et al. 2003). As a result, at the frequency where the curvature of the input spectrum is negative, a smaller number of photons are present around that frequency with respect to the case of the standard CMB spectrum (where the spectral curvature, in brightness temperature units, is zero), and the resulting SZE-21cm is smaller than the standard one; on the other hand, where the curvature is positive a larger number of photons is present and the SZE-21cm is higher than the standard one. 

We also find that the minimum point in the input radiation spectrum ($\nu\sim70$ MHz) corresponds to a maximum point in the SZE-21cm; this is due to the fact that a minimum point in the input spectrum means a smaller number of photons with respect to the standard CMB: as a consequence, when subtracting the input spectrum to calculate the SZE-21cm (see eq.\ref{eq.Deltaxmod}), the resulting emission is stronger than for the standard SZE. The opposite behaviour is observed at the frequencies where the input radiation spectrum has its maximum points ($\nu\sim 45$ and 120 MHz), that are close to the minimum points of the SZE-21cm; in this case, the correspondence is less precise with respect to the previous case because the maximum points in the input spectrum are less sharped than the minimum one, and the convolution of photons with those at surrounding frequencies produces a slight shift in the frequency of the minimum points in the SZE-21cm.  

In the following we discuss more detailed and new results obtained for the specific case of the SZE-21cm produced by i) thermal electron populations, that provide the dominant contribution to the SZE observed in galaxy clusters, and by ii) non-thermal electrons populations, that are present in clusters that show non-thermal activity (i.e. radio halos or relics) and in the extended lobes of radiogalaxies. This can be done by using the corresponding functions $f_e(p)$ in eq.(\ref{eq.p1s}), i.e. a maxwellian distribution for a thermal population and a power-law distribution for a non-thermal population. A specific analysis on the relevance of relativistic effects in the SZE-21cm is also presented.

\section{The SZE-21cm: detailed spectral analysis}

In the following we discuss first our results obtained for the benchmark modified background radiation scenario (solid line in Fig. \ref{cmb_modified}), and then for the set of other  modified radiation background models shown in  Fig.\ref{cmb_modified}. 

\begin{figure}[ht]
\begin{center}
%\vspace{-10cm}
{
 \epsfig{file=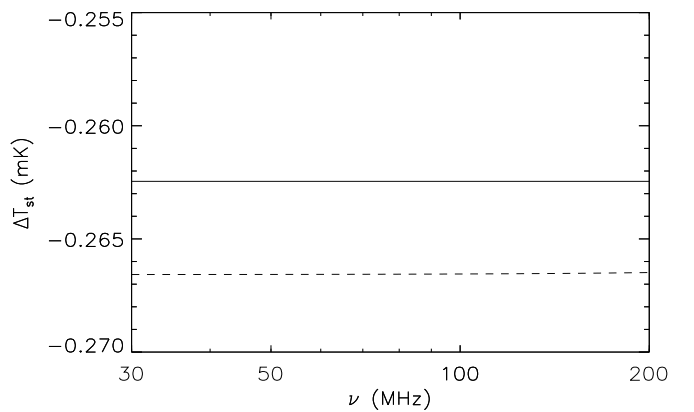,height=6.cm,width=9.cm,angle=0.0}
}
\end{center}
 \caption{\footnotesize{The standard SZE (in units of brightness temperature relative to the CMB) 
for a thermal plasma ($kT=5$ keV and $\tau=5\times10^{-3}$; solid line) and for a non-thermal plasma
($s=3.5$, $p_1=10$ and $\tau=1\times10^{-4}$; dashed line). The SZE is shown in the radio frequency range where the 21-cm radiation background are visible.
 }}
 \label{sz_dtemp}
\end{figure}

We start our discussion, for the sake of clarity, by showing the spectral shape of the standard SZE, $\Delta T_{st}$, for the unmodified CMB spectrum. 
Figure \ref{sz_dtemp} shows the standard SZE, not modified by the 21-cm line radiation field, in units of brigthness temperature relative to the CMB for the case of a galaxy cluster with thermal plasma at temperature $kT=5$ keV and with optical depth $\tau=5\times10^{-3}$, and for the case of a non-thermal plasma with a single power-law spectrum $N(p) \sim p^{-s}$ for $p\geq p_1$, with $s=3.5$, $p_1=10$ and 
$\tau=1\times10^{-4}$.
We notice that the standard SZE is a constant line in units of CMB brightness temperature in the Rayleigh-Jeans (RJ) regime ($h \nu \ll k T_{CMB}$) for both the case of a thermal SZE 
%while it is very slightly modified at $\nu \simgt 0.1$ GHz  in 
and for the case of a non-thermal, relativistic plasma typical of the radiogalaxy lobes (we assume here a steep spectrum $S_{\nu} \propto \nu^{-\alpha_R}$ with $\alpha_R = (s-1)/2 =1.25$) 

\begin{figure}[ht]
\begin{center}
%\vspace{-10cm}
{
 \epsfig{file=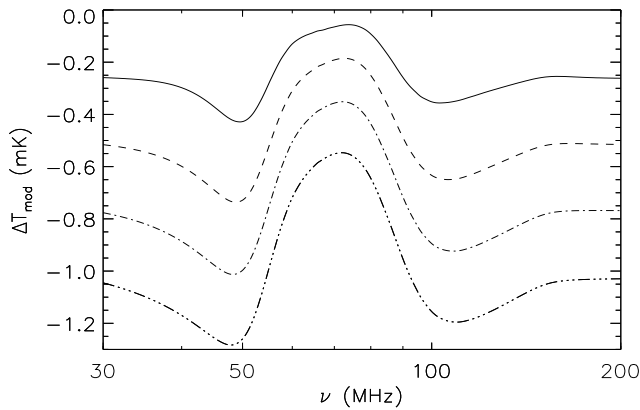,height=6.cm,width=9.cm,angle=0.0}
}
\end{center}
 \caption{\footnotesize{The SZE-21cm (in units of brightness temperature relative to the CMB) 
for thermal plasma at temperature $kT=5$, 10, 15 and 20 keV, shown by the solid, dashed, dot-dashed and dash-three dots lines, respectively. A constant value $\tau=5\times10^{-3}$ has been used in the calculations.
 }}
 \label{sz_therm}
\end{figure}

The thermal SZE-21cm, $\Delta T_{mod}$, is shown in Fig. \ref{sz_therm} for the case of thermal plasma in galaxy clusters for four different electron temperatures of 5, 10, 15 and 20 keV. We find that the spectral shape of the thermal SZE-21cm changes for different electron temperatures, consistently with the effects of relativistic corrections that are fully considered in our approach, while its amplitude increases with the cluster temperature, which is consistent with the notion that the SZE amplitude increases with increasing the cluster Compton parameter $y= {\sigma_T \over m_e c^2} \int  d\ell P_e$, that reads $y  \propto kT \cdot \tau$ for the case of a thermal intracluster medium (Colafrancesco et al. 2003).

\begin{figure}[ht]
\begin{center}
%\vspace{-10cm}
{
  \epsfig{file=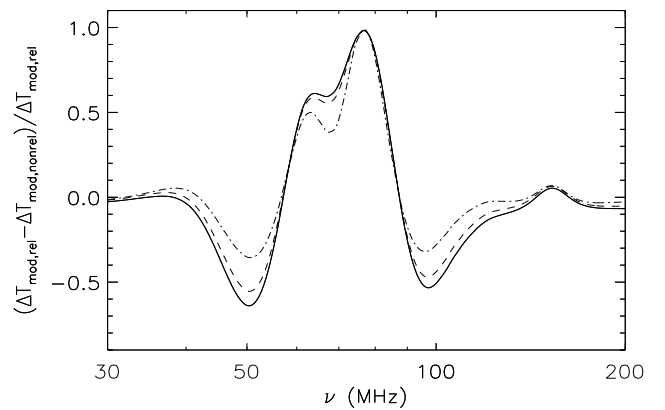,height=6.cm,width=9.cm,angle=0.0}
}
\end{center}
 \caption{\footnotesize{Percentage difference between the relativistic result and the non-relativistic one for the SZE-21cm for galaxy clusters with temperatures of 20 keV (solid line), 15 keV (dashed line) and 7 keV (dot-dashed line).}
}
 \label{sz_nonrel}
\end{figure}

We verified  the level of the error  done when using a non-relativistic approach to calculate the SZE-21cm w.r.t. our full relativistic approach. In Fig.\ref{sz_nonrel} we show the percentage difference between the results of the two calculations for clusters with electron temperatures of 20, 15 and 7 keV calculated with the relativistic and the non-relativistic approaches (as in the case of Cooray 2006). We find that the percentage difference is different from 0 (i.e., the case in which the non-relativistic calculation gives the same result than the relativistic one) at almost all frequencies. 
As discussed in details in the Appendix, we also note that the percentage difference has local maxima (in absolute value)  in correspondence of the points where the second derivative of the input spectrum has its maxima and minima, i.e. at $\nu\sim50$, 60, 77 and 95 MHz (see lower panel in Fig.\ref{app_fig1}). This is related, as discussed for the shape of the SZE-21cm, to the fact that the SZE is produced by a convolution of the input spectrum photon distribution with photons at surrounding frequencies. The non-relativistic calculation considers a shape of the function $P(s)$ which is narrower than the one in the relativistically correct calculation (see, e.g., Birkinshaw 1999, Colafrancesco et al. 2003). Therefore, when the curvature (positive or negative) of the input radiation spectrum is maximum, the error done by convolving the input spectrum with a function $P(s)$ narrower than the correct one is larger, because it implies to lose the contribution from the photons with farther frequencies. As a consequence, the more the input spectrum is different from a straight line, the larger is the error done by using the non-relativistic calculation. In the Appendix we expand these considerations by discussing also the other three input models considered for the input radiation spectrum we use in this paper.

For the case of a cluster with a temperature of 20 keV, the percentage difference reaches at its local maxima/minima values of the order of $\approx 65 \%$, $\approx 60 \%$, $\approx 100 \%$ and $\approx 50 \%$ at frequencies $\nu \approx 50, 60, 77, 95$ MHz, respectively, which introduce therefore substantial modifications in the value of the SZE-21cm calculated in the non-relativistic approach.
For the other temperatures, the percentage error is smaller, but still of the order of at least $30\%$ at the previous frequencies, and at $\sim77$ MHz the percentage error is $\sim100\%$ independently on the cluster temperature.  
For this reason we conclude that in order to perform a correct study of the SZE-21cm it is mandatory to use the full relativistic formalism as described in our paper.

\begin{figure}[ht]
\begin{center}
%\vspace{-10cm}
{
 \epsfig{file=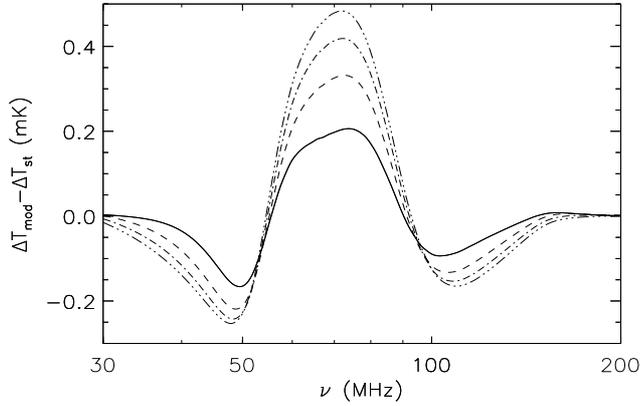,height=6.cm,width=9.cm,angle=0.0}
}
\end{center}
 \caption{\footnotesize{Difference between the SZE-21cm and the standard SZE (in units of brightness temperature relative to the CMB) for thermal plasma at temperature $kT=5$, 10, 15 and 20 keV, shown by the solid, dashed, do-dashed and dash-three dots lines, respectively, as in Fig.\ref{sz_therm}. A constant value $\tau=5\times10^{-3}$ has been used in the calculations.
 }}
 \label{sz_therm_diff}
\end{figure}

In Figure \ref{sz_therm_diff} we show the difference between the value of $\Delta T$ for the thermal SZE-21cm and the standard thermal SZE on the unmodified CMB  (note that this last SZE is a constant value for all frequencies in the considered range, as shown in Fig.\ref{sz_dtemp}) in order to highlight the spectral difference between the two effects and between the thermal effects calculated for different electron temperatures.
We notice that the main differences appear around 50 MHz and in the range $\approx 60-80$ MHz (reflecting the Ly$\alpha$ spin coupling effect), and in the range $100-150$ MHz (reflecting the UV ionization effect during the EoR).

\begin{figure}[ht]
\begin{center}
%\vspace{-10cm}
{
 \epsfig{file=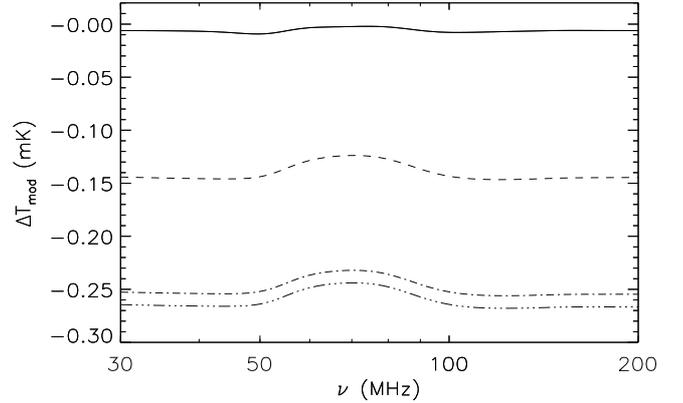,height=6.cm,width=9.cm,angle=0.0}
}
\end{center}
 \caption{\footnotesize{The SZE-21cm (in units of brightness temperature relative to the CMB) 
for non-thermal electrons with a power-law spectrum with $s=3.5$ and $p_1=0.1$, 1, 5 and 10, 
shown by the solid, dashed, dot-dashed and dash-three dots lines, respectively. 
A constant value $\tau=1\times10^{-4}$ has been used in the calculations.
 }}
 \label{sz_lp}
\end{figure}

\begin{figure}[ht]
\begin{center}
%\vspace{-10cm}
{
 \epsfig{file=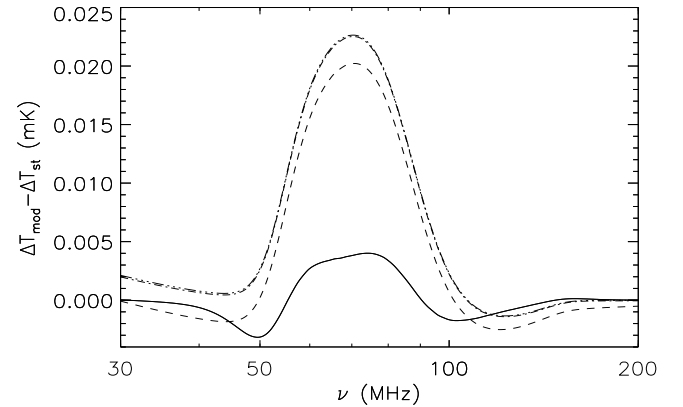,height=6.cm,width=9.cm,angle=0.0}
}
\end{center}
 \caption{\footnotesize{Difference between the SZE-21cm and the standard SZE 
(in units of brightness temperature relative to the CMB) 
for non-thermal electrons with a power-law spectrum with $s=3.5$ and $p_1=0.1$, 1, 5 and 10, 
shown by the solid, dashed, dot-dashed and dash-three dots lines, respectively. 
A constant value $\tau=1\times10^{-4}$ has been used in the calculations.
 }}
 \label{sz_lp_diff}
\end{figure}

To investigate the non-thermal SZE-21cm  effect produced by non-thermal (or relativistic) electrons residing, e.g., in the radiogalaxy lobes or in cluster radio halos/relics, we consider an electron population with a single power-law spectrum with index  $s=3.5$ and various values of the minimum electron momentum $p_1$. Figure \ref{sz_lp} shows the non-thermal SZE-21cm for values $p_1=0.1$, 1, 5 and 10. 
The non-thermal SZE-21cm has an amplitude that increases (in modulus) with increasing values of $p_1$,
for a constant value of $\tau$.
We show in  Fig. \ref{sz_lp_diff} the difference between the non-thermal SZE-21cm and the standard non-thermal SZE where the CMB spectrum is not modified. 
The largest differences of the non-thermal SZE-21cm with respect to the standard one take place at frequencies similar to the thermal case, and the differences with the thermal case are more important for high values of $p_1$, i.e. when the scattering electrons are more energetic.

\begin{figure}[ht]
\begin{center}
%\vspace{-10cm}
{
 \epsfig{file=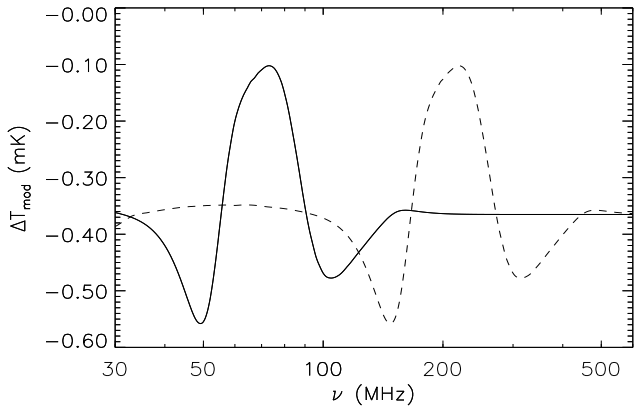,height=6.cm,width=9.cm,angle=0.0}
}
\end{center}
 \caption{\footnotesize{For an illustrative description of the possible redshift-dependence of the overall modified background model, we show the SZE-21cm (in units of brightness temperature relative to the CMB)  for a thermal plasma at temperature $kT=7$ keV and with $\tau=5\times10^{-3}$ for the modified CMB spectrum with the values of $z$ taken from the original model (solid line), and for a modified CMB spectrum  in which all components are globally shifted in frequency by a factor 3 (dashed line).
 }}
 \label{sz_7kev_z2}
\end{figure}

We also check how the shape of the SZE-21cm depends on the frequency of the modifications to the overall radiation field, that depends on the assumed redshift range in which the various mechanisms operating during the DA and EoR act to modify the original CMB spectrum. To this purpose, for an illustrative description of the possible redshift-dependence of the overall modified background model, we show the frequency shape of the resulting SZE-21cm when the redshift of the input modified radiation field is varied.
To this aim, we use a typical galaxy cluster with a thermal electron plasma at a temperature of 7 keV and optical depth 
$\tau=5\times10^{-3}$, and we compare the total SZE-21cm as previously discussed with 
the one in which the background spectrum is shifted globally in frequency by a factor 3 (see Fig. \ref{sz_7kev_z2}). 
With this illustrative example, we are considering the possibility that the redshifts at which the various phenomena (e.g., collisions, Ly-$\alpha$ interactions, UV ionization) can be different from the ones assumed in the benchmark model. Thus, from the frequency at which the different effects in the SZE-21 cm are observed, it is possible to derive the redshift at which these effects took place, and in principle  determine the full cosmic history of the DA and EoR.
%
%
% \begin{figure}[ht]
% \begin{center}
%\vspace{-10cm}
% {
%  \epsfig{file=sz21cm_7keV_z2col.ps,height=6.cm,width=9.cm,angle=0.0}
%  \epsfig{file=sz21cm_15keV_z2col.ps,height=6.cm,width=9.cm,angle=0.0}
% }
% \end{center}
%  \caption{\footnotesize{\textbf{[Not changed]} The SZE-21cm (in units of brightness temperature relative to the CMB) 
% for a thermal plasma at temperature $kT=7$ keV (upper panel) and 15 keV (lower panel)
% and with $\tau=5\times10^{-3}$ for a modified CMB spectrum 
% in which the components are shifted in frequency by a factor 3 %$(1+z)$, with $z=2$ 
% for the absorption/collision term (solid line), and %with $z=0$ 
% are not modified for the Ly-$\alpha$ (dashed line) and
% the X-ray heating (dot-dashed line) terms.
%  }}
%  \label{sz_7kev_z2col}
% \end{figure}
%

%\subsection{Detailed 21-cm scenarios}

\begin{figure}[ht]
\begin{center}
%\vspace{-10cm}
{
 \epsfig{file=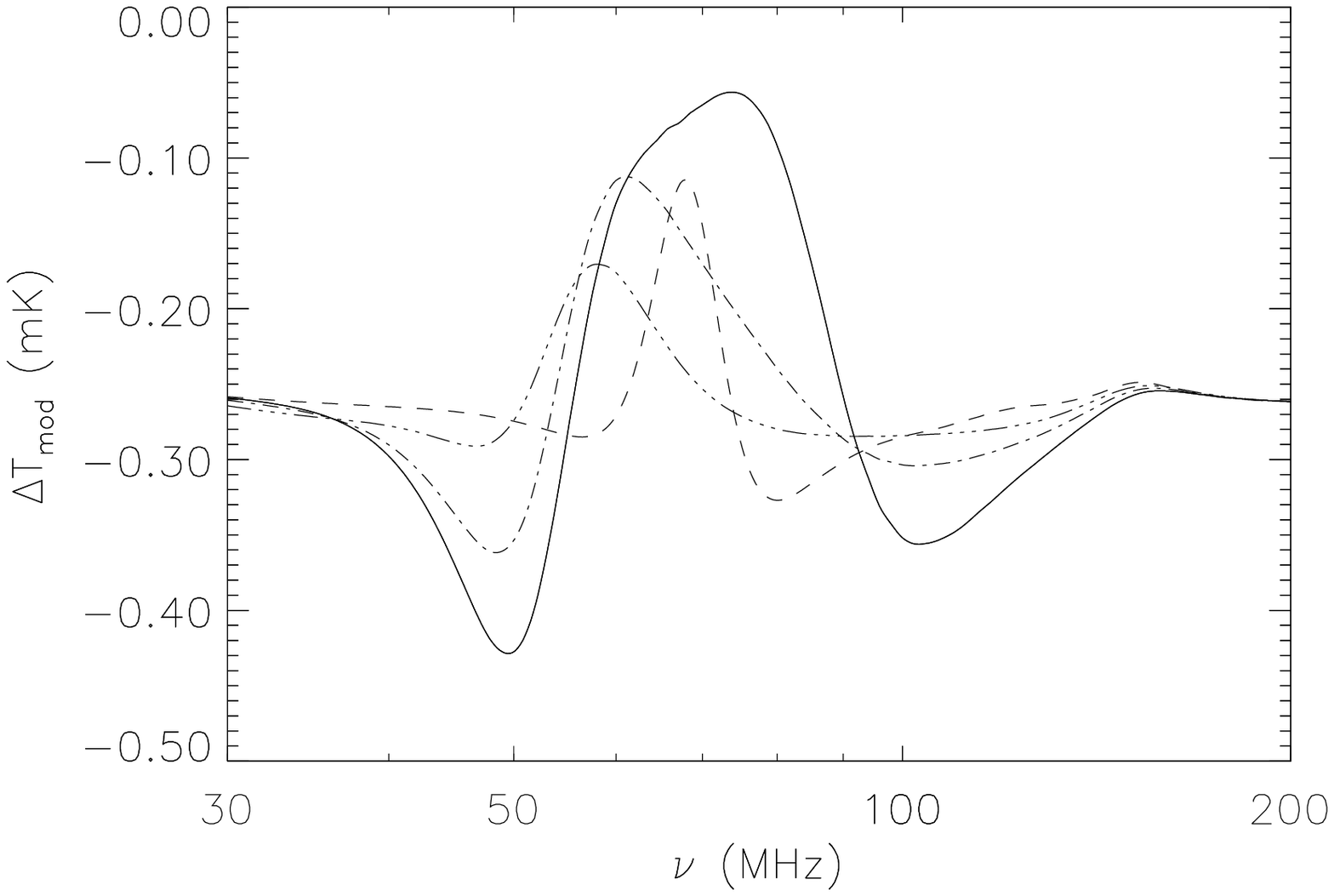,height=6.cm,width=9.cm,angle=0.0}
 \epsfig{file=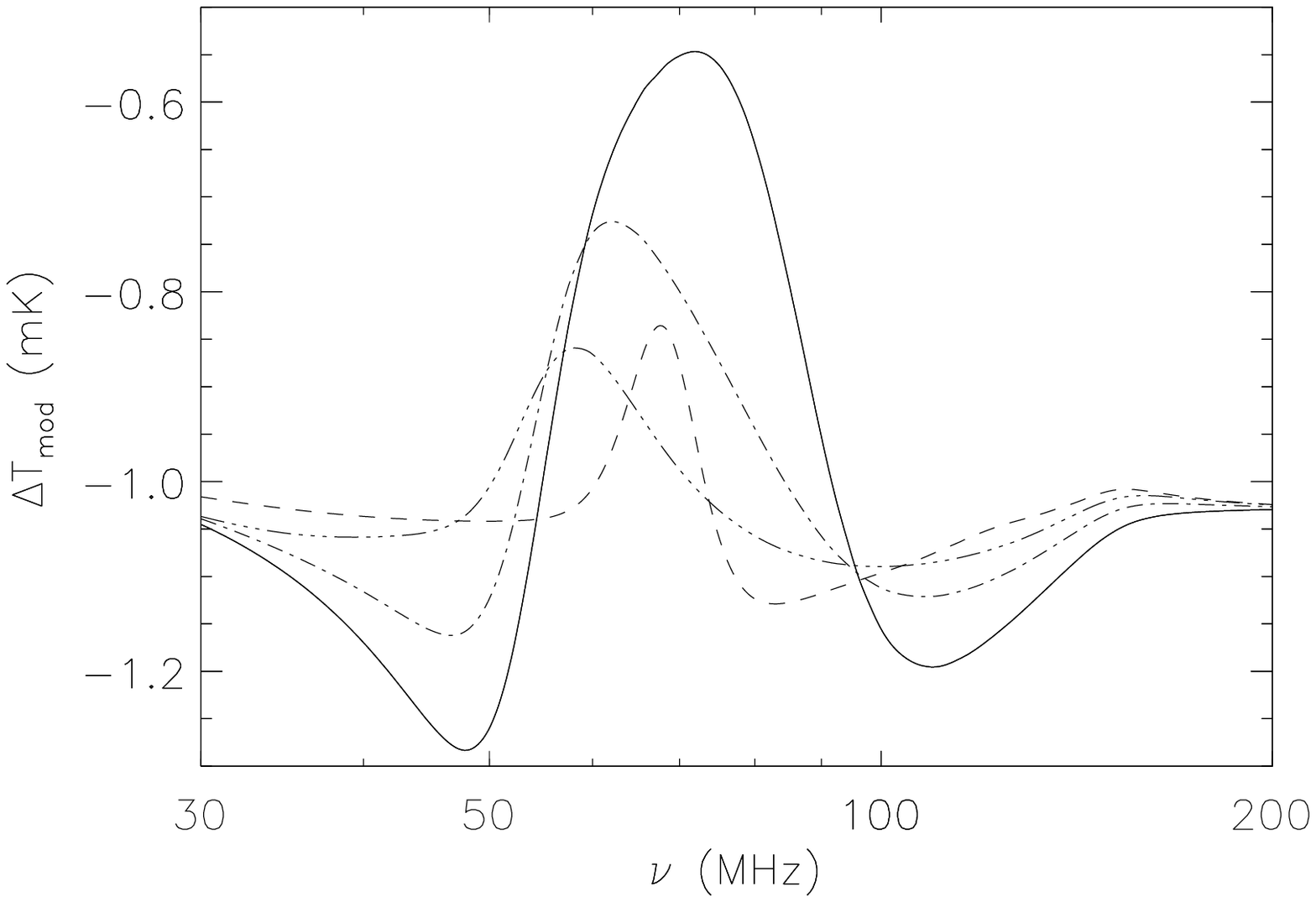,height=6.cm,width=9.cm,angle=0.0}
}
\end{center}
 \caption{\footnotesize{The SZE-21cm (in units of brightness temperature relative to the CMB) 
for a thermal plasma at temperature $kT=5$ keV (upper panel) and 20 keV (lower panel)
and with $\tau=5\times10^{-3}$ for a modified CMB spectrum with a fiducial model without Dark Matter (solid line), an extreme model without Dark Matter (dashed line), a fiducial model with Dark Matter with $M_{min}=10^{-3}$ M$_\odot$ (dot-dashed line), and a fiducial model with Dark Matter with $M_{min}=10^{-6}$ M$_\odot$ (three dots-dashed line).
 }}
 \label{sz_4models}
\end{figure}

\begin{figure}[ht]
\begin{center}
%\vspace{-10cm}
{
 \epsfig{file=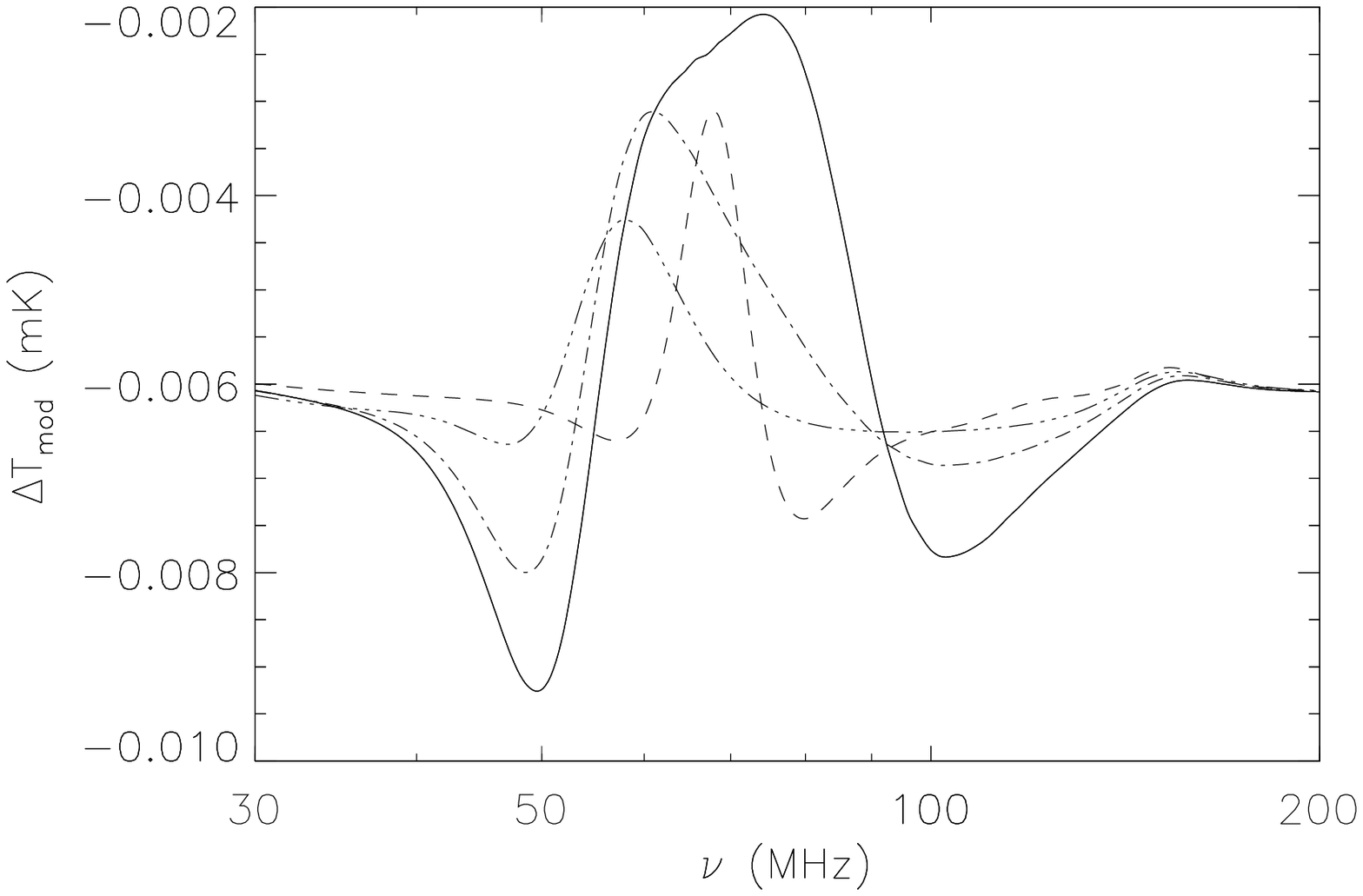,height=6.cm,width=9.cm,angle=0.0}
 \epsfig{file=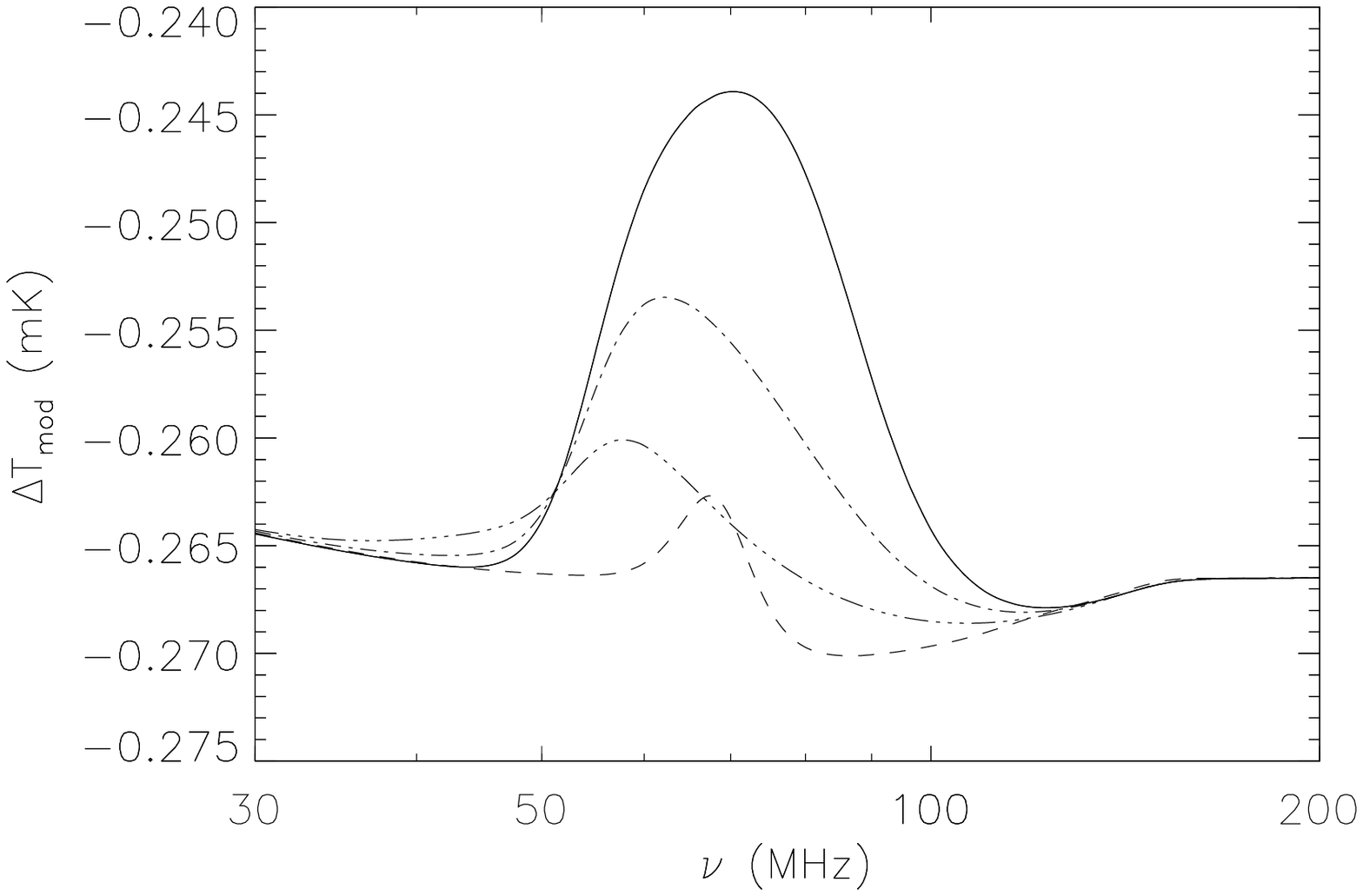,height=6.cm,width=9.cm,angle=0.0}
}
\end{center}
 \caption{\footnotesize{The SZE-21cm (in units of brightness temperature relative to the CMB) 
for a non-thermal plasma with $s=3.5$ and $p_1=0.1$ (upper panel) and 10 (lower panel) and with $\tau=1\times10^{-4}$
for a modified CMB spectrum with a fiducial model without Dark Matter (solid line),
an extreme model without Dark Matter (dashed line), 
a fiducial model with Dark Matter with $M_{min}=10^{-3}$ M$_\odot$ (dot-dashed line), and
a fiducial model with Dark Matter with $M_{min}=10^{-6}$ M$_\odot$ (three dots-dashed line).
 }}
 \label{sz_4models_lp}
\end{figure}

By using the other models of the modified radiation background described in Sect. 2.2, we obtained the results shown in Figure \ref{sz_4models}, where the thermal SZE-21cm spectrum for clusters with 5 and 20 keV is plotted, and in Figure \ref{sz_4models_lp}, where instead the non-thermal SZE-21cm with $s=3.5$ and $p_1=0.1$ and 10 is plotted.
As we can see, while the spectral shape of the non-thermal SZE-21 cm is very similar to the thermal one for $p_1=0.1$, for high values of $p_1$ the main difference is the damping of the features produced by the Ly-$\alpha$ spin coupling effect at $\sim60$ and 100 MHz. The effect of considering a higher heating rate, both from usual astrophysical sources and from DM, is to increase the temperature of the IGM, to which the spin temperature is linked by the Ly-$\alpha$ coupling, and as a result the peak in the SZE-21cm in the 60--80 MHz frequency range is damped, with different spectral shapes depending on the Dark Matter properties.

These results therefore show that the SZE-21cm can be also considered as a tool to probe both the amount of DM in the universe and the minimal mass of DM halos collapsed at early epochs. The DM abundance can be probed using the amplitude and the spectral shape of the SZE-21cm in two best frequency ranges: around $\sim 50$ MHz and at $\approx 60-90$ MHz, where the sensitivity to the DM density is higher. 
The sensitivity to $M_{min}$ for the DM halos is best achievable at $\nu \approx 50-70$ MHz where the effect of $M_{min}$ increases the amplitude of the SZE-21cm and shifts its maximum in frequency.

\section{Discussion}

In the full relativistic description of the SZE-21cm we found that the following properties are important for the correct use of this technique:

i) The scattering properties of high-energy electrons need a full relativistic treatment: avoiding this will generate percentage differences up to about 100 \% at the relevant frequencies where this effect can be observed. This is ensured in our approach through a self-consistent computation of the SZE-21cm. 

ii) We find that the amplitude of the SZE-21cm and its variations  w.r.t. the standard SZE (using the non-modified CMB spectrum) are larger for clusters with high temperature (see Fig.\ref{sz_therm_diff}) and for non-thermal electron plasmas with high values of the minimum momentum $p_1$ (see Fig.\ref{sz_lp_diff}), i.e. when the high-energy electrons are more important.

iii) Studying the detailed spectrum of the SZE-21cm allows to derive precise information on the epochs at which the CMB has been modified and on the physical mechanism that produced such modifications during the DA and EoR (see Fig.\ref{sz_7kev_z2}).

iv) The thermal and non-thermal SZE-21cm have peculiar spectral shapes (see Figs. \ref{sz_therm_diff}, \ref{sz_lp_diff} and \ref{sz_4models}-\ref{sz_4models_lp}). Thus, it is possible, in principle, to derive information also on the existence and the properties of the electron population in cosmic structures also from very low-$\nu$ observations of the SZE. We note that this property is complementary with the results of previous studies, in accordance with which the properties of non-thermal electrons can be derived from the study of the SZE at high frequencies (see, e.g., Colafrancesco, Marchegiani \& Buonanno 2011 for the case of the Bullet Cluster).

\subsection{Differential analysis technique and foreground contamination}

Observations of the SZE-21cm can be carried out with radio interferometers since the modification associated with low-redshift scattering can be established from differential observations towards and away from galaxy clusters and other cosmic structures containing diffuse thermal and non-thermal plasmas.
Unlike an experiment to directly establish the cosmic 21-cm frequency spectrum at low radio frequencies involving a total intensity measurement of the sky, the differential observations with a radio interferometer are less affected by issues such as the exact calibration of the observed intensity using an external source, and the confusion from galactic foregrounds that are uniform over angular scales larger than a typical cluster, such as the Galactic synchrotron background at low radio frequencies. Also, since the SZE does not depend on redshift, it is more suitable to study sources located at large distances, allowing to reduce the importance of the cluster radio emissions (both diffuse and point-like sources) with respect to the SZE, and allowing to detect a larger number of sources, thus increasing the possibility to obtain more precise results by studying this effect in many sources at cosmological scales.\\
The resulting modification to the 21-cm spectrum due to the thermal SZE-21cm is expected at the level of a few tenths mK brightness temperature relative to the CMB. Therefore, such a small modification challenges an easy detection, but for upcoming radio interferometers (like the SKA), the specific spectral signatures would allow to produce a relatively clean detection. In addition, multi-object SZE-21cm observations could be facilitated by the fact that the instantaneous field-of-view of upcoming interferometers is expected to be more than 100 square degrees and one expects to detect simultaneously hundreds, or more, massive clusters in such wide fields.\\ 
Therefore, the SZE-21cm effect can be effectively used  to establish the global features in the mean 21-cm spectrum generated during and prior to the EoR. 
We note that it is also possible to produce cluster population studies with the SZE-21cm (e.g., cluster counts and redshift distribution) and use them as cosmological probes. These goals make desirable to build a technique allowing to study a large number of objects (including galaxy clusters in merging and relaxed states, radio halo and cooling flow clusters, radio galaxy lobes), and to study objects at high redshift.

\begin{figure}[ht]
\begin{center}
%\vspace{-10cm}
{
 \epsfig{file=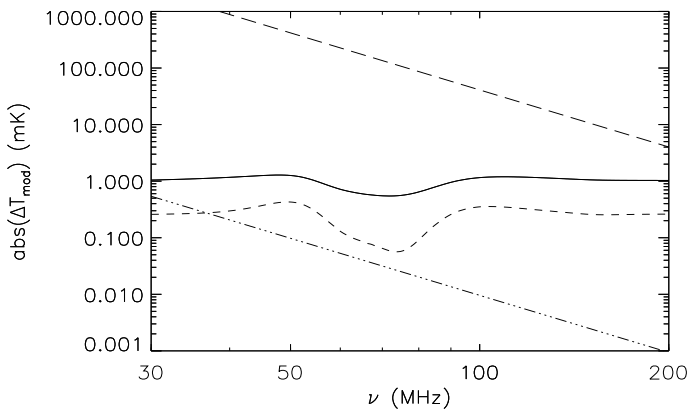,height=6.cm,width=9.cm,angle=0.0}
}
\end{center}
 \caption{\footnotesize{The SZE-21cm (in units of Brightness Temperature relative to the CMB and in absolute value) 
for a thermal plasma with temperature $kT=20$ keV (solid line) and 5 keV (dashed line), and with $\tau=5\times10^{-3}$, compared with a spectrum similar to that of Coma radio halo (long-dashed line), and with the same spectrum for a Coma-like cluster located at $z=1$.
 }}
 \label{sz_synchr}
\end{figure}

Even if the differential measurements of the SZE-21cm avoid contamination from foreground/background emissions on scales larger than the cluster/radiogalaxy size,  another possible source of contamination is the synchrotron radio emission within galaxy clusters and radio galaxies lobes. This contamination should decrease for objects at large distances, because the synchrotron emission varies with the luminosity distance as $D_L^{-2}$, whereas the SZE does not vary with the distance of the source. For nearby objects, the synchrotron emission at low frequencies can be much stronger than the SZE. In Fig. \ref{sz_synchr} we show a comparison between two cases of the SZE-21 cm (for thermal plasma with temperature of 5 and 20 keV and optical depth $\tau=5\times10^{-3}$), a spectrum similar to that of the Coma radio halo (approximated as a perfect power law), and a spectrum of a Coma-like cluster located at $z=1$. We note the at all frequencies we are interested, the synchrotron emission is much larger than the SZE for a nearby cluster like Coma; so, it is necessary to study the cluster radio halo spectrum to separate the two contributions.
At higher-$z$, however, the radio halo flux decreases rapidly while the SZE-21cm remains unchanged thus providing a lower level of contamination and an easier subtraction procedure.\\
Another possible source of contamination is given by the point radio sources in galaxy clusters; in this case, the goal to separate this contribution from the SZE-21 cm signal is easier, since it is possible to use both the spectral information we have at other frequencies and the spatial information, in order to remove the contribution from point sources.

\subsection{Detectability with SKA}

We discuss now the detectability of the SZE-21 cm with the SKA1-low instrument. We extracted the performance of SKA1-low from  the SKA1 System Baseline Design document (see Dewdney et al. 2012).

\begin{figure}[ht]
\begin{center}
%\vspace{-10cm}
{
  \epsfig{file=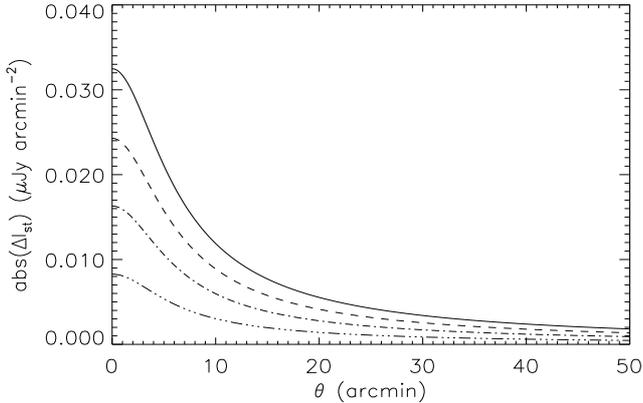,height=6.cm,width=9.cm,angle=0.0}
}
\end{center}
 \caption{\footnotesize{Surface brightness profile of the standard SZE in absolute value
for thermal plasma with temperatures $kT=20$ (solid line), 15 (dashed), 10 (dot-dashed) and 5 (three dots-dashed) keV, and calculated for $\tau_0=5\times10^{-3}$, $\theta_c=300$ arcesc, $\beta=0.75$, $\theta_{max}=10\theta_c$.
 }}
 \label{sz_bright}
\end{figure}

First of all, we calculate the loss of signal at small angular radii produced by the finite extension of the interferometer. For this purpose, we calculate the SZE flux from an isothermal cluster with a gas density profile given by a $\beta$-profile:
\begin{equation}
n_e(r)=n_{e,0}\left[1+ \left( \frac{r}{r_c} \right )^2 \right ]^{-\frac{3}{2}\beta}
\end{equation}
(Cavaliere \& Fusco-Femiano 1976). For such a cluster, the optical depth at a projected distance $\theta$ from the center of the cluster is given by the expression:
\begin{equation}
\tau(\theta)=\tau_0 \left [1+\left(\frac{\theta}{\theta_c}\right)^2\right]^{\frac{1}{2}-\frac{3}{2}\beta}
\label{tau_profile}
\end{equation}
(Colafrancesco et al. 2003), where $\theta_c=r_c/D_A$ and $D_A$ is the angular diameter distance of the cluster. We assume $\tau_0=5\times10^{-3}$, $\beta=0.75$, $\theta_c=300$ arcsec and calculate the flux up to an angular size $\theta_{max}=10\theta_c$.\\
The reference spatial resolution of SKA1-low at 110 MHz, corresponding to a minimum baseline of 50 km, is $\theta_{min}\sim11$ arcsec. Since at first order in $\tau$ the SZE-21cm is proportional to the product of the SZE spectral function and of the cluster optical depth (see, e.g., Colafrancesco et al. 2003), we can estimate that the lack of sensitivity for angular scales $\theta < \theta_{min}$ is given by the ratio between the optical depth integrated in this small $\theta$ range and the total one, and it implies a signal loss of the order of
\begin{equation}
\frac{\int_0^{\theta_{min}} 2\pi \theta \tau(\theta) d\theta}{\int_0^{\theta_{max}} 2\pi \theta \tau(\theta) d\theta}
\sim1.1\times10^{-4} .
\end{equation}
To have an idea about the intensity of the signal we should expect, we plot in Fig. \ref{sz_bright}
the surface brightness profiles of the standard SZE at the frequency of 110 MHz for the optical depth profiles in the eq.(\ref{tau_profile}), with the same parameters values described above, and for the temperatures of 20, 15, 10, and 5 keV. Therefore, in the inner part (e.g., within a radius of $\sim20$ arcmin) of a galaxy cluster with high temperature we can estimate a SZE signal of the order of  $\sim10$ $\mu$Jy and, as a consequence, the loss of signal due to the finite baseline configuration of the SKA1 is of the order of $\sim$ nJy, and therefore does not affect our results.

% To study the detectability of the SZE-21 cm signal, we compare the flux calculated for the 21 cm modified CMB and the one calculated for the non-modified CMB with the sensitivity of SKA1-low for 100 kHz bandwith, 1000 hrs of integration, 2 polarizations, no taper, no weight.
% We show the result in Figure \ref{sz_ska}. This figure shows that the SZE-21 cm can be detected \textbf{with 1000 hrs integration time} at frequencies $\nu\simgt70$ MHz for clusters with very high temperature ($K_BT=20$ keV) and at $\nu\simgt90$ MHz for low temperature clusters ($K_BT=5$ keV).

% \begin{figure}[ht]
% \begin{center}
%\vspace{-10cm}
% {
%   \epsfig{file=sz21_ska.ps,height=6.cm,width=9.cm,angle=0.0}
% }
% \end{center}
%  \caption{\footnotesize{The spectra of the fluxes of the SZE-21cm (in units of $\mu$Jy and in absolute value with the solid lines) and the SZE for a non-modified CMB (dashed lines)
% for thermal plasma with temperatures $kT=20$ (green), 15 (black), 10 (red) and 5 (cyan) keV, and calculated for $\tau_0=5\times10^{-3}$, $\theta_c=300$ arcesc, $\beta=0.75$, $\theta_{max}=10\theta_c$, compared with the SKA1-low sensitivity for 100 kHz bandwith, 1000 hrs of integration, 2 polarizations, no taper, no weight (thick line).
%  }}
%  \label{sz_ska}
% \end{figure}

\begin{figure}[ht]
\begin{center}
%\vspace{-10cm}
{
  \epsfig{file=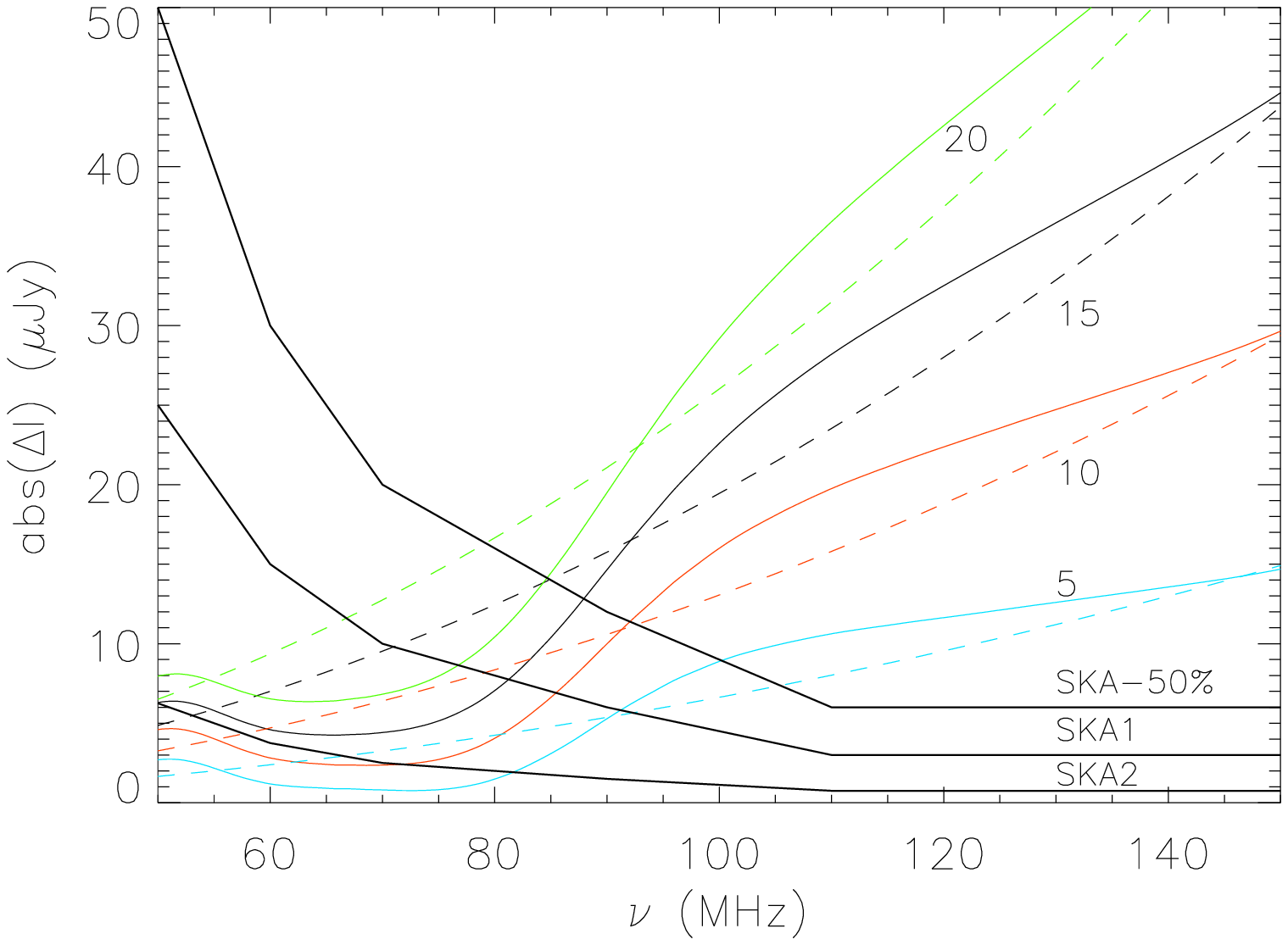,height=6.cm,width=9.cm,angle=0.0}
  \epsfig{file=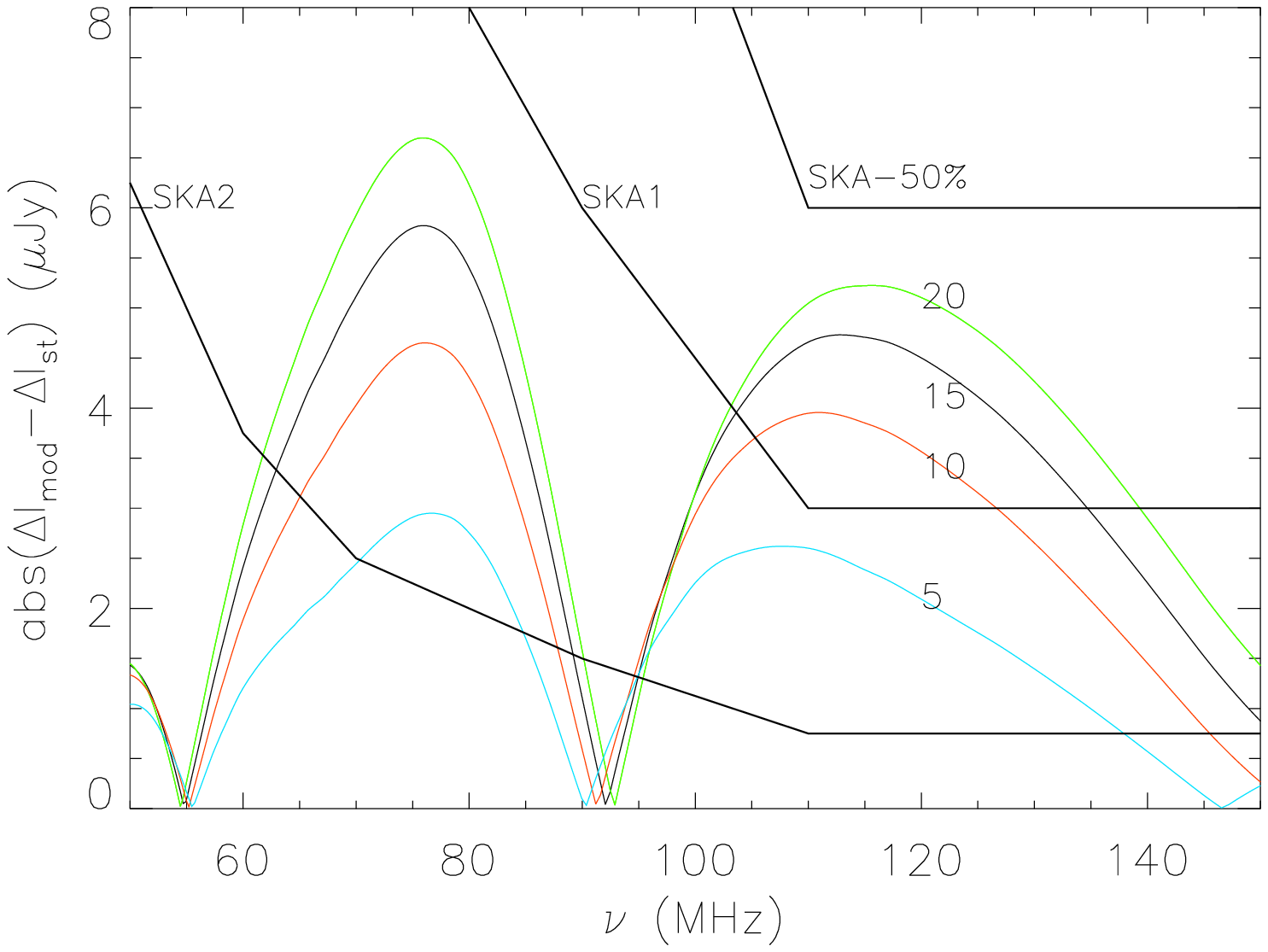,height=6.cm,width=9.cm,angle=0.0}
}
\end{center}
 \caption{\footnotesize{Upper panel: the spectra of the fluxes of the SZE-21cm $\Delta I_{mod}$ (in units of $\mu$Jy and in absolute value with the solid lines) and the SZE for a non-modified CMB $\Delta I_{st}$ (dashed lines).
Lower panel: the absolute value of the difference between the SZE-21cm and the standard SZE for a non-modified CMB.
Both panels are using for the modified CMB the fiducial model without Dark Matter (solid line in Fig.\ref{cmb_modified}).
Both panels are for thermal plasma with temperatures $kT=20$ (green), 15 (black), 10 (red) and 5 (cyan) keV, and calculated for $\tau_0=5\times10^{-3}$, $\theta_c=300$ arcesc, $\beta=0.75$, $\theta_{max}=10\theta_c$, compared with the SKA-50\%, SKA1-low, and SKA 2 sensitivities for 100 kHz bandwith, 1000 hrs of integration, 2 polarizations, no taper, no weight (thick lines).
 }}
 \label{sz_ska2}
\end{figure}

\begin{figure}[ht]
\begin{center}
%\vspace{-10cm}
{
  \epsfig{file=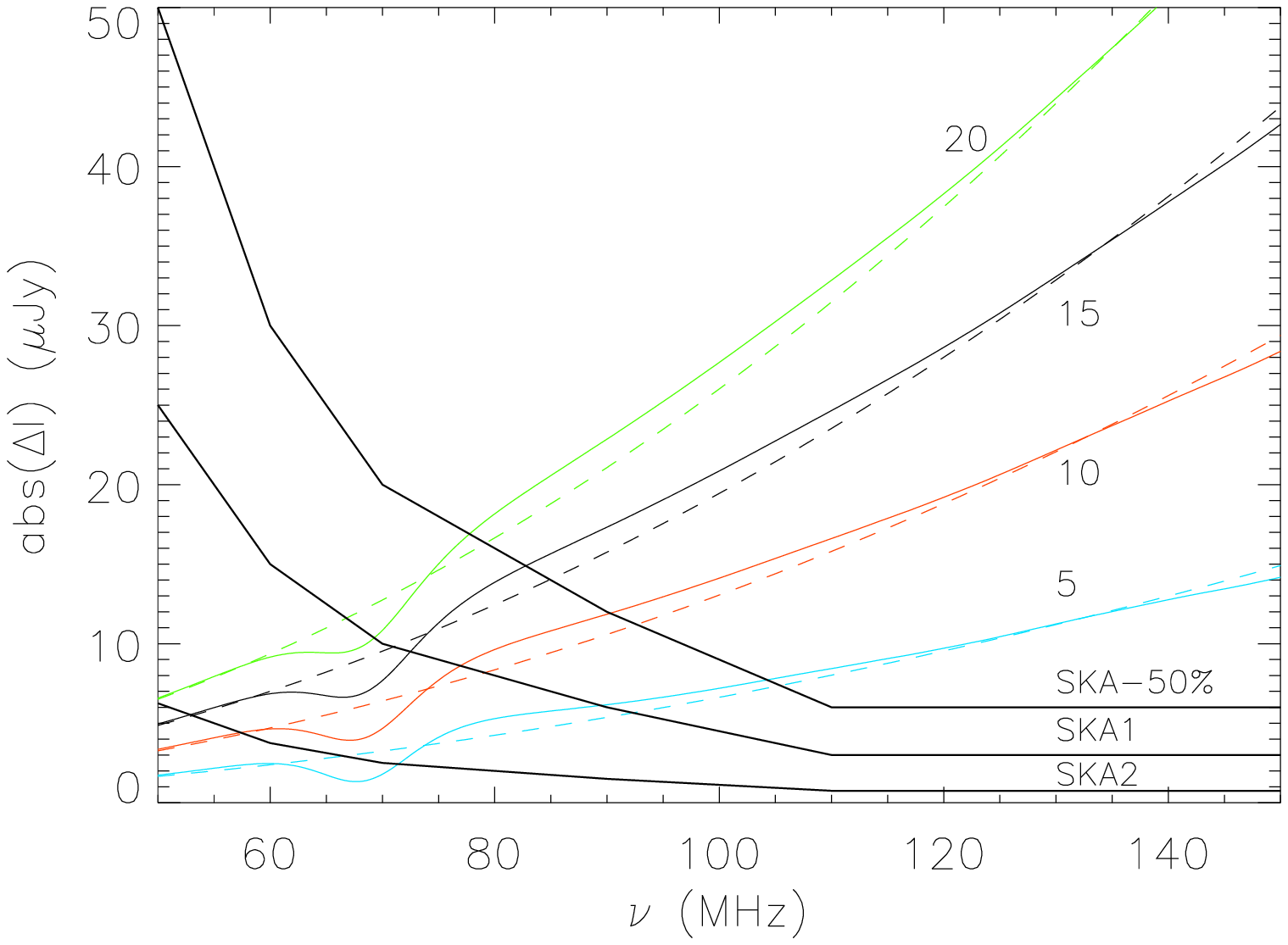,height=6.cm,width=9.cm,angle=0.0}
  \epsfig{file=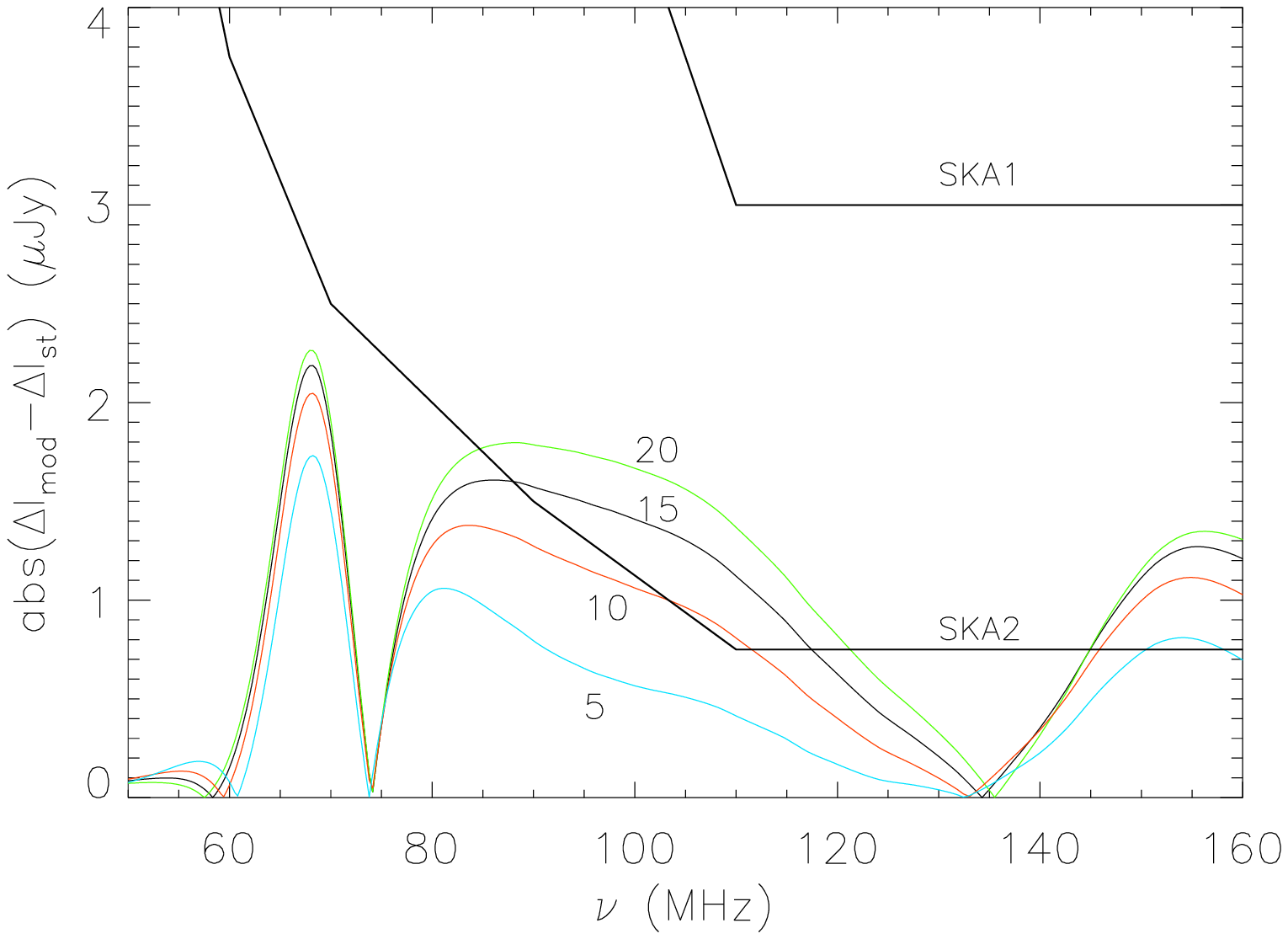,height=6.cm,width=9.cm,angle=0.0}
}
\end{center}
 \caption{\footnotesize{Like Fig.\ref{sz_ska2} but for an extreme model without Dark Matter for the modified CMB (dashed line in Fig.\ref{cmb_modified}).
 }}
 \label{sz_ska2_dt2}
\end{figure}

\begin{figure}[ht]
\begin{center}
%\vspace{-10cm}
{
  \epsfig{file=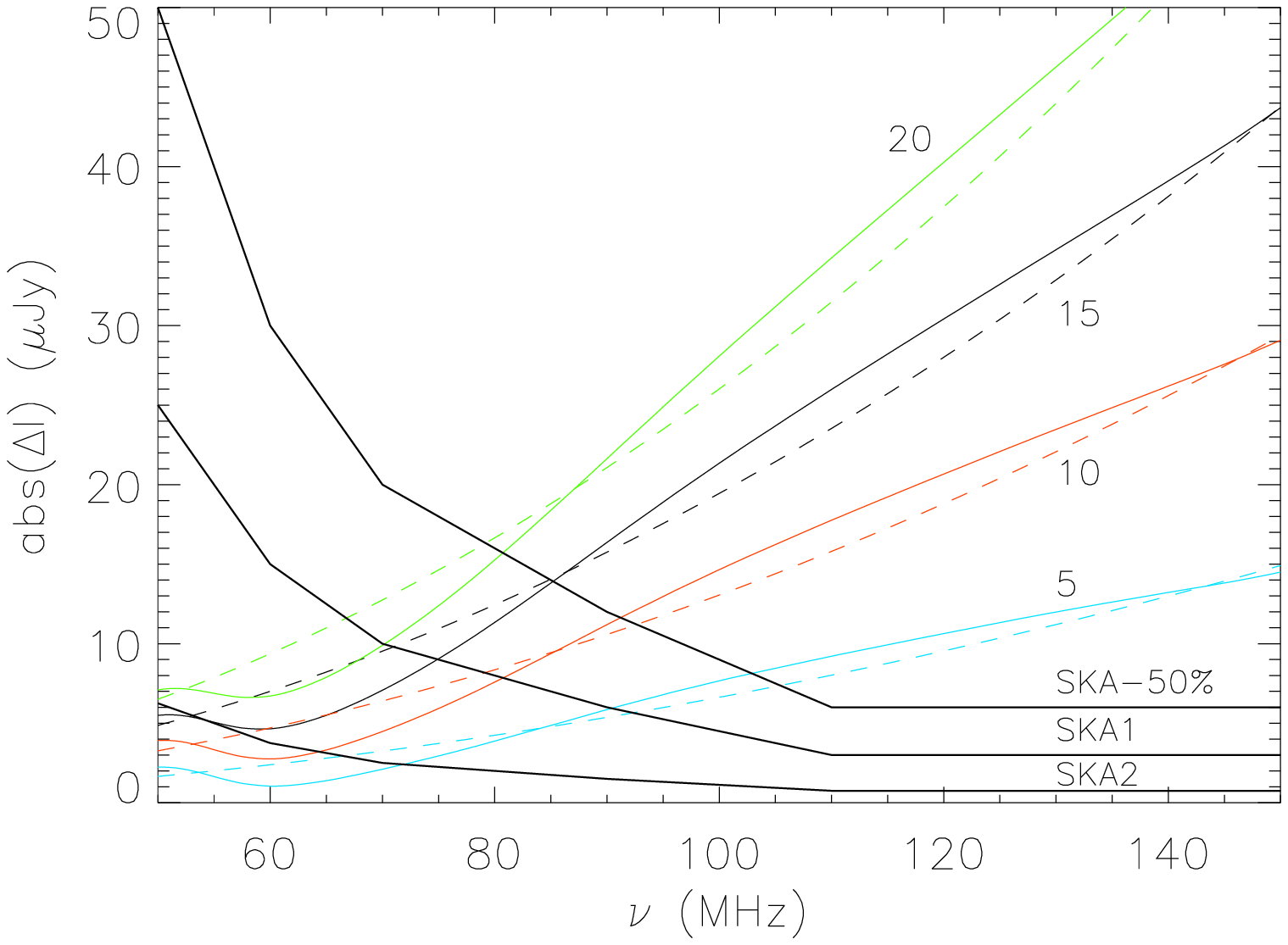,height=6.cm,width=9.cm,angle=0.0}
  \epsfig{file=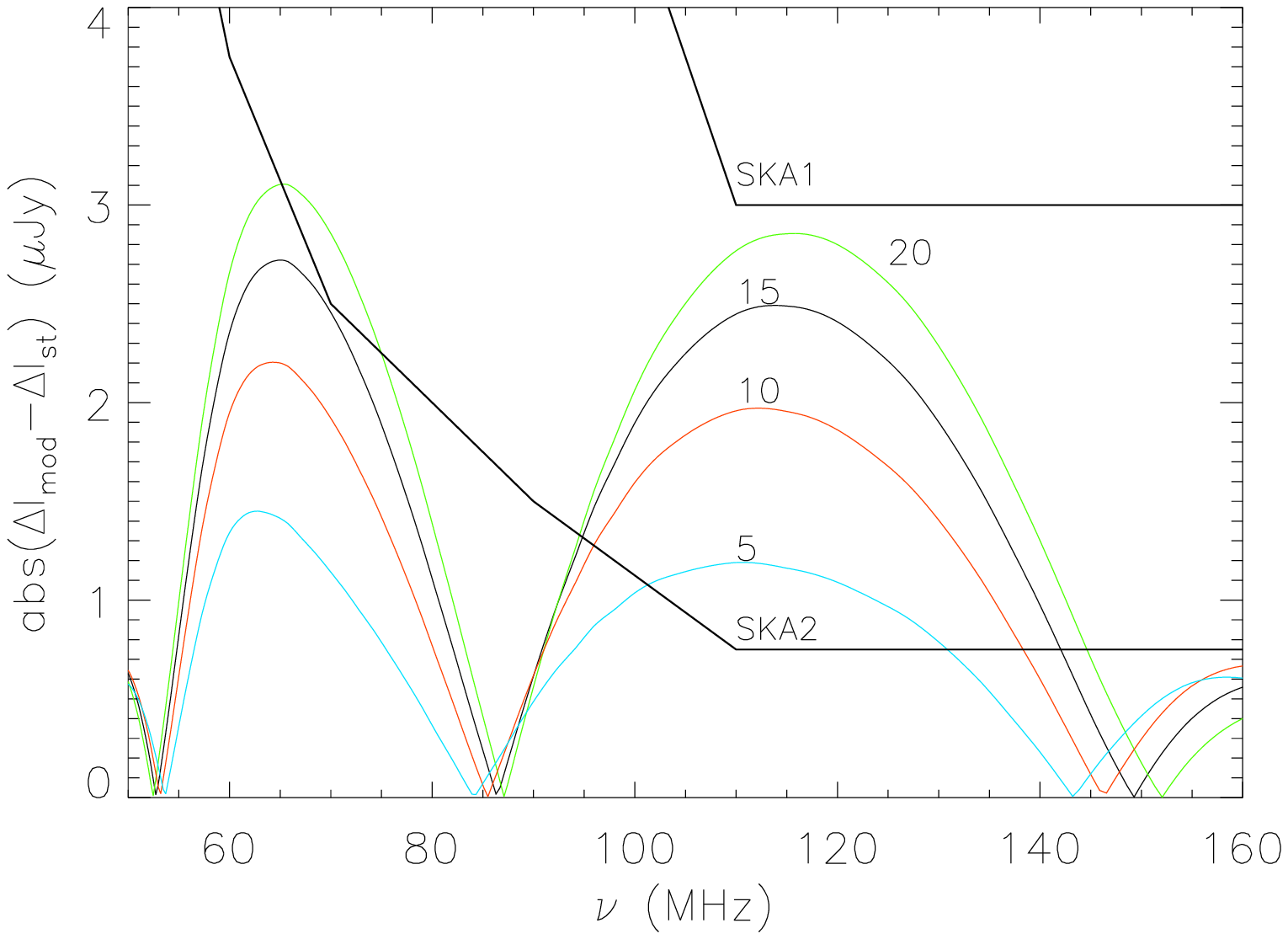,height=6.cm,width=9.cm,angle=0.0}
}
\end{center}
 \caption{\footnotesize{Like Fig.\ref{sz_ska2} but for a fiducial model with Dark Matter with $M_{min}=10^{-3}$ M$_\odot$ for the modified CMB (dot-dashed line in Fig.\ref{cmb_modified}).
 }}
 \label{sz_ska2_dt3}
\end{figure}

\begin{figure}[ht]
\begin{center}
%\vspace{-10cm}
{
  \epsfig{file=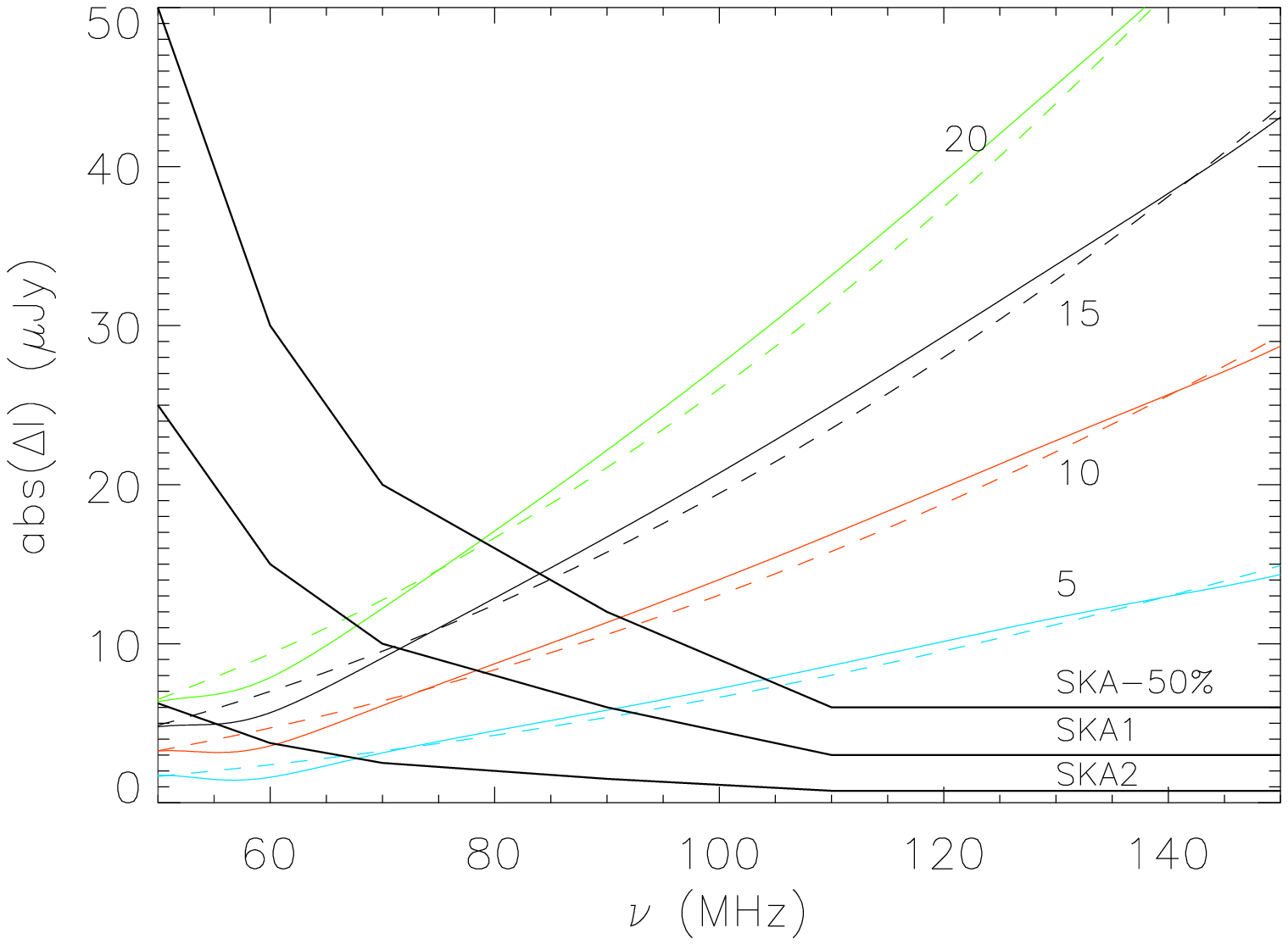,height=6.cm,width=9.cm,angle=0.0}
  \epsfig{file=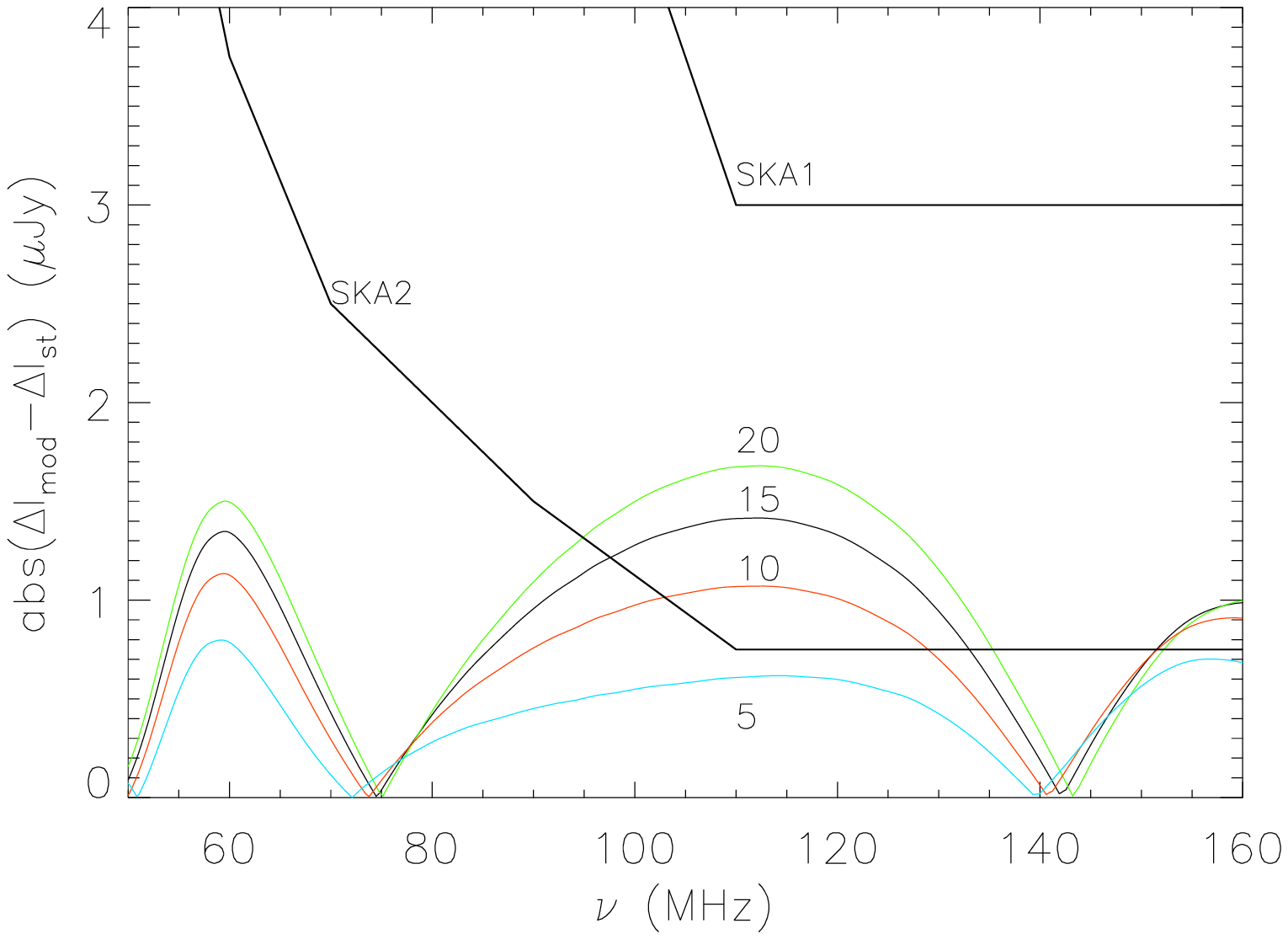,height=6.cm,width=9.cm,angle=0.0}
}
\end{center}
 \caption{\footnotesize{Like Fig.\ref{sz_ska2} but for a fiducial model with Dark Matter with $M_{min}=10^{-6}$ M$_\odot$ for the modified CMB (three dots-dashed line in Fig.\ref{cmb_modified}).
 }}
 \label{sz_ska2_dt4}
\end{figure}

To study the detectability of the SZE-21 cm signal, we compare the flux calculated for the modified CMB spectrum, $\Delta I_{mod}$, and the one calculated for the non-modified CMB spectrum, $\Delta I_{st}$, with the sensitivities of SKA-50\%, SKA1, and SKA2 for 100 kHz bandwith, 1000 hrs of integration, 2 polarizations, no taper, no weight.
We show the result in Figures \ref{sz_ska2}-\ref{sz_ska2_dt4} for the different radiation background models we use in this paper. 

For our benchmark model, the SZE-21cm can be detected with SKA1-low with 1000 hrs integration time at frequencies $\nu\simgt75$ MHz for clusters with very high temperature ($kT=20$ keV) and at $\nu\simgt90$ MHz for low temperature clusters ($kT=5$ keV). With SKA-50\% the SZE-21cm can be detected at higher frequencies (85 and 100 MHz for hot and cold clusters respectively), and with SKA2 it can be detected at small frequencies (50 and 80 MHz), giving the possibility to study the EoR and the Dark Ages until very high redshift ($z\sim30$).

The possibility to discriminate between the SZE-21 cm and the standard SZE signals is more challenging: the difference between the two signals is always at most of the order of few $\mu$Jy (see lower panel in Figure \ref{sz_ska2}), so it requires to measure the signal with high precision, and at frequencies where the differences are larger. Good frequency channels for this purpose can be found at $\sim$75 MHz (where the SZE-21cm is lower than the standard SZE because of the Ly-$\alpha$ coupling effect), and at 100--110 MHz (where the SZE-21cm is stronger because of the UV ionization effect). 
Because of the better sensitivity of SKA1-low at its high frequency band, the best frequency range where we can obtain information on the SZE-21 cm is $\nu \simgt 100$ MHz. However, also in this frequency range the difference between the two signals is of the order of $\mu$Jy, so very deep observations, and very accurate data analysis procedures are required for this purpose, together with the fact that it is necessary to use clusters with high values of electron temperature and optical depth.\\ 
We further show  that with SKA2 the difference between the modified and the standard SZE can be detected at frequency $\nu \simgt60$ MHz in galaxy clusters with temperature $kT\simgt15$ keV 
and at $\simgt65$ MHz in clusters with temperature $kT\simgt10$ keV for an integration time of 1000 hrs. 

For the other models we use, detecting the difference between the modified and the standard SZE is more challenging.
In general, it is not possible to detect this difference with SKA1; only in the case of the model with Dark Matter with $M_{min}=10^{-3}$ M$_\odot$ it would be possible detect this difference by increasing the integration time by a factor $\sim3$ for the hottest clusters at a frequency around 110 MHz. With SKA2, the detection is possible at frequencies 85--120 MHz (only for cluster temperatures $kT>10$ keV) and $\simgt145$ MHz for the case of extreme heating without Dark Matter, at 65--75 MHz (only for $kT\sim20$ keV) and 95--145 MHz in the case of the model with Dark Matter with $M_{min}=10^{-3}$ M$_\odot$, and at 95--135 MHz and $\simgt150$ MHz (for $kT\simgt10$ keV) in the case of the model with Dark Matter with $M_{min}=10^{-6}$ M$_\odot$.

A promising strategy can be designed to study the SZE at higher frequencies (with experiments like, e.g., SPT, ACT, Millimetron) in order to derive precise information on the parameters of the ICM, and then use these constraints to obtain a better estimate of the properties of the SZE-21 cm with SKA1-low and SKA2.

% \begin{figure}[ht]
% \begin{center}
% \vspace{-10cm}
% {
%   \epsfig{file=sz21_ska_diff_v3.ps,height=6.cm,width=9.cm,angle=0.0}
% }
% \end{center}
%  \caption{\footnotesize{The absolute value of the difference between the SZE-21cm and the standard SZE for a non-modified CMB for thermal plasma with temperatures $kT=20$ (green), 15 (black), 10 (red) and 5 (cyan) keV, and calculated for $\tau_0=5\times10^{-3}$, $\theta_c=300$ arcesc, $\beta=0.75$, $\theta_{max}=10\theta_c$, compared with the SKA1-low sensitivity for 100 kHz bandwith, 10000 (thick solid line) and 20000 (thick dashed line) hrs of integration, 2 polarizations, no taper, no weight.
%  \textbf{Paolo: check the integration times and the sensitivity curves for SKA-low, SKA 2 and SKA 50\%}
%  }}
%  \label{sz_ska_diff}
% \end{figure}

% \begin{figure}[ht]
% \begin{center}
% \vspace{-10cm}
% {
%   \epsfig{file=sz21_ska2_diff_v3_rev.ps,height=6.cm,width=9.cm,angle=0.0}
% }
% \end{center}
%  \caption{\footnotesize{The absolute value of the difference between the SZE-21cm and the standard SZE for a non-modified CMB for thermal plasma with temperatures $kT=20$ (green), 15 (black), 10 (red) and 5 (cyan) keV, and calculated for $\tau_0=5\times10^{-3}$, $\theta_c=300$ arcesc, $\beta=0.75$, $\theta_{max}=10\theta_c$, compared with the \textbf{SKA-50\%, SKA1-low, and SKA 2 sensitivities for 100 kHz bandwith, 1000 hrs of integration, 2 polarizations, no taper, no weight (thick lines)}.
% \textbf{Paolo: check the integration times and the sensitivity curves for SKA-low, SKA 2 and SKA 50\%}
%  }}
%  \label{sz_ska2_diff}
% \end{figure}

% \begin{table*}[htb]{}
% \vspace{2cm}
% \begin{center}
% \begin{tabular}{|*{9}{c|}}
% \hline 
% Frequency  & $\Delta I_{mod}$ & $\Delta I_{mod}-\Delta I_{st}$ & $\Delta I_{mod}$ & $\Delta I_{mod}-\Delta I_{st}$ &
% $\Delta I_{mod}$ & $\Delta I_{mod}-\Delta I_{st}$ & $\Delta I_{mod}$ & $\Delta I_{mod}-\Delta I_{st}$  \\
%    MHz    &  $\mu$Jy & $\mu$Jy & $\mu$Jy & $\mu$Jy & $\mu$Jy & $\mu$Jy & $\mu$Jy & $\mu$Jy \\
% \hline 
%  & \multicolumn{2}{c|} {5 keV} & \multicolumn{2}{c|} {10 keV} & \multicolumn{2}{c|} {15 keV} & \multicolumn{2}{c|} {20 keV} \\
% \hline
% 62 & -1.0 & 1.5 & -2.6 & 2.4 & -4.4 & 3.1 & -6.4 & 3.6 \\ 
% 72 & -0.77 & 2.7 & -2.4 & 4.3 & -4.6 & 5.5 & -7.2 & 6.3 \\
% 96 & -7.7 & -1.6 & -14 & -1.8 & -20 & -1.8 & -26 & -1.6 \\
% 110 & -11 & -2.6 & -20 & -3.9 & -28 & -4.7 & -37 & -5.0 \\
%  \hline
%  \end{tabular}
%  \end{center}
%  \caption{\footnotesize{\textbf{Modified} SZE-21 cm fluxes ($\Delta I_{mod}$) and differences between between the \textbf{modified and the standard SZE} calculated under the same assumptions used in the Figure \ref{sz_ska2} and at the frequencies indicated in the first column.
%  }}
%  \label{tab.fluxes}
%  \end{table*}

\section{Conclusions}

The goal of obtaining information on the physical processes occurred during the DA and EoR by measuring the SZE-21 cm with SKA is challenging, but possible if pursued with good theoretical and observational strategies. 

Observations have to be carried out towards high temperature and high optical depth clusters to maximize both the overall signal and the difference between the standard and the modified SZE. The best frequency ranges of observation of the SZE-21cm are between $\sim$ 90 and 120 MHz, where the difference between the standard and the modified SZE is maximum. 
In our benchmark model, the sensitivity of SKA1-low is good enough to detect this difference with 1000 hours of integration, 
whereas for the other background models the difference between the standard and the modified SZE can be detected only with SKA2 for the same integration time in frequencies bands that depend on the background model and the temperature of the cluster.

Together with very deep observations, a very accurate theoretical analysis is required, where the full formalism to calculate the SZE and detailed models for describing the effect of the cosmological 21-cm background on the CMB spectrum have to be used. In addition, we find that a very important strategy will be the detailed study of the SZE at higher frequencies in order to estimate the gas parameters to be used as prior constraints for the study of the SZE-21 cm at low frequencies.\\
Observations in the frequency bands of SKA1-mid are also very important to disentangle the SZE from the cluster synchrotron emission. In this respect, the use of high-redshift clusters can alleviate the problem, since the radio emission decreases as $D_L^{-2}$, whereas the SZE is not depending on the cluster distance.

The detection of the non-thermal SZE-21 cm appears to be more challenging, since the signal is much fainter with respect to the thermal one, especially regarding the difference between the standard and the modified SZE, that can be also a factor of $\sim 10^2$ smaller with respect to the thermal case. 
However, the different spectral features can allow, in principle, a detection of this signal and hence an estimate of non-thermal cluster properties independently of measurements in other spectral bands. We note here that it is possible to strategize the search of this signal in objects where the non-thermal components are dominant, such as in the case of radio galaxies lobes. In this case, objects with more energetic electrons (i.e. with harder radio spectra), large optical depth (for which a good indication could be a strong radio luminosity) and high redshift are preferable.

The independence of the SZE from the redshift can allow the study of the SZE-21cm in a larger number of objects spread over a wider redshift range, therefore producing statistical studies aimed at maximizing the detectable signal, and detect the properties of the 21-cm background and of the early DM halos over a large set of spatial directions, allowing in such a way a better understanding of the full cosmic history of the physical processes  occurring in the Dark Ages  and the Epoch of Reionization.

\begin{acknowledgements}
S.C. acknowledges support by the South African Research Chairs Initiative of the Department of Science and Technology and National Research Foundation and by the Square Kilometre Array (SKA). P.M. and M.S.E. acknowledge support from the DST/NRF SKA post-graduate bursary initiative. We thank C. Evoli for providing the numerical files of the models in Fig.1, and A. Ferrara, M. Birkinshaw and A. Tailor for useful discussions. We thank the Referee for several useful comments and suggestions that allowed us to improve the presentation of our results.
\end{acknowledgements}

{}

\appendix

\section{The relation between the error done by using the non-relativistic calculation and the properties of the 
input radiation field.}

In this Appendix, we estimate the error done when the SZE-21cm is calculated by using a non-relativistic approach, as a function of the properties of the input spectrum, using the four models shown in Fig.\ref{cmb_modified}.

For the standard SZE the input radiation is a Planck black-body spectrum which at low frequency has a constant brightness temperature, and the resulting SZE $\Delta T_{st}$ is a constant as well (see Fig.\ref{sz_dtemp}). It is important to note that the Planck spectrum is a smooth function, and we noticed that because of this smoothness the difference between the use of a relativistic approach and a non-relativistic approach in computing the SZE is smaller for low electron temperatures and at low frequencies (see, e.g., Colafrancesco et al. 2003).
However, when computing the SZE-21 cm,  the shape of the input radiation spectrum plays an important role for the determination of the error done in the calculation of the SZE-21cm using a non-relativistic approach.

\begin{figure}[ht]
\begin{center}
%\vspace{-10cm}
{
 \epsfig{file=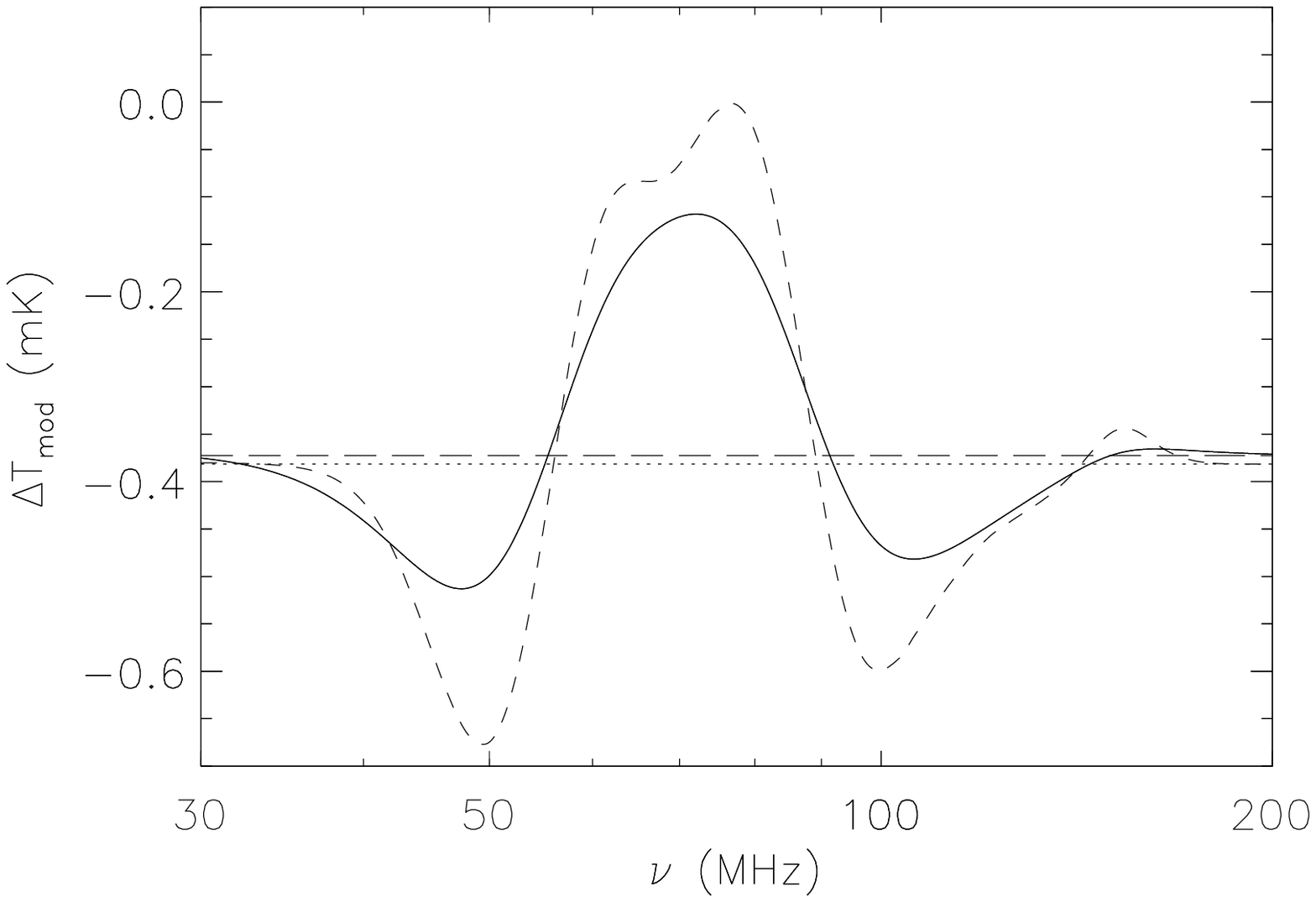,height=6.cm,width=9.cm,angle=0.0}
 \epsfig{file=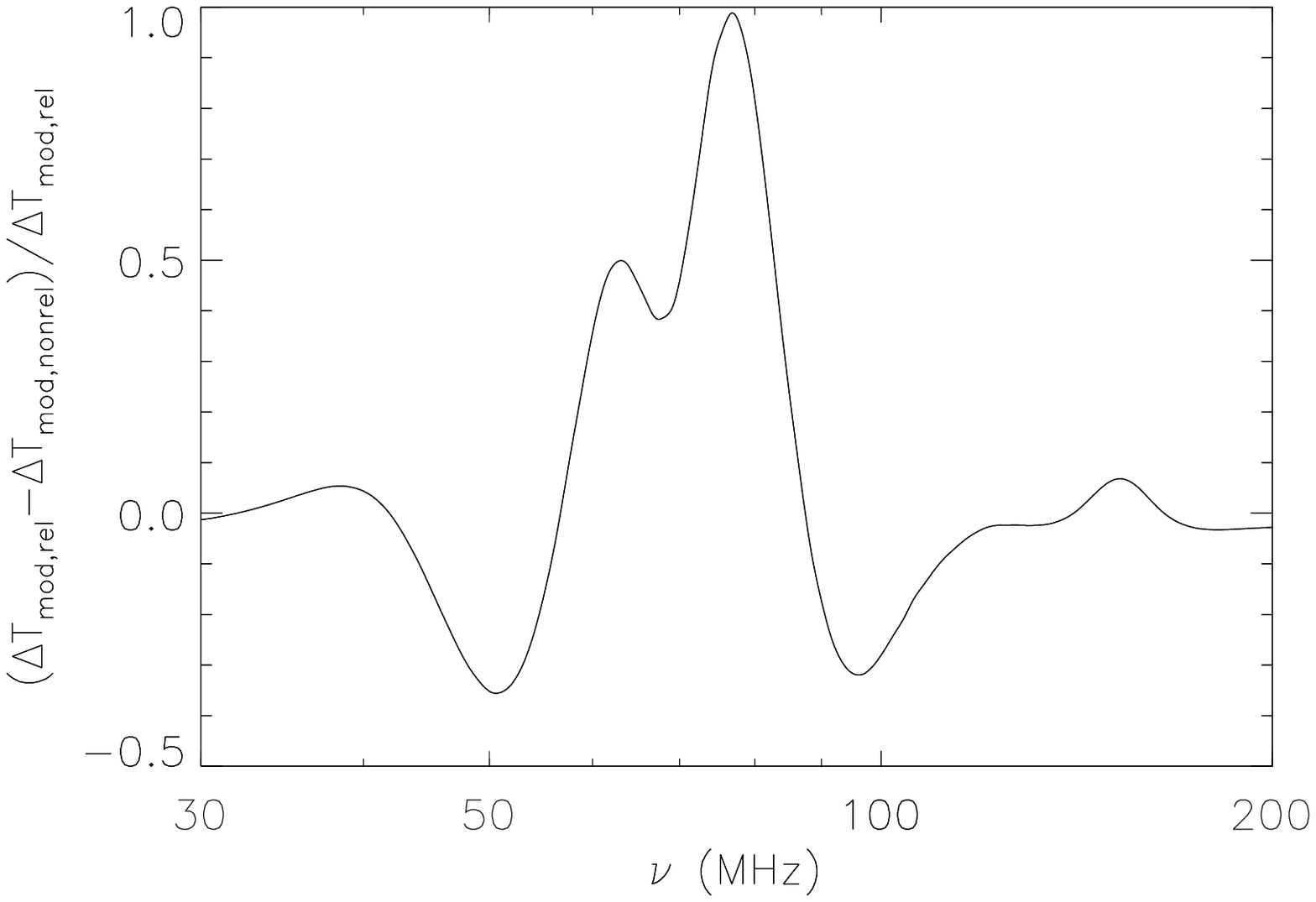,height=6.cm,width=9.cm,angle=0.0}
 \epsfig{file=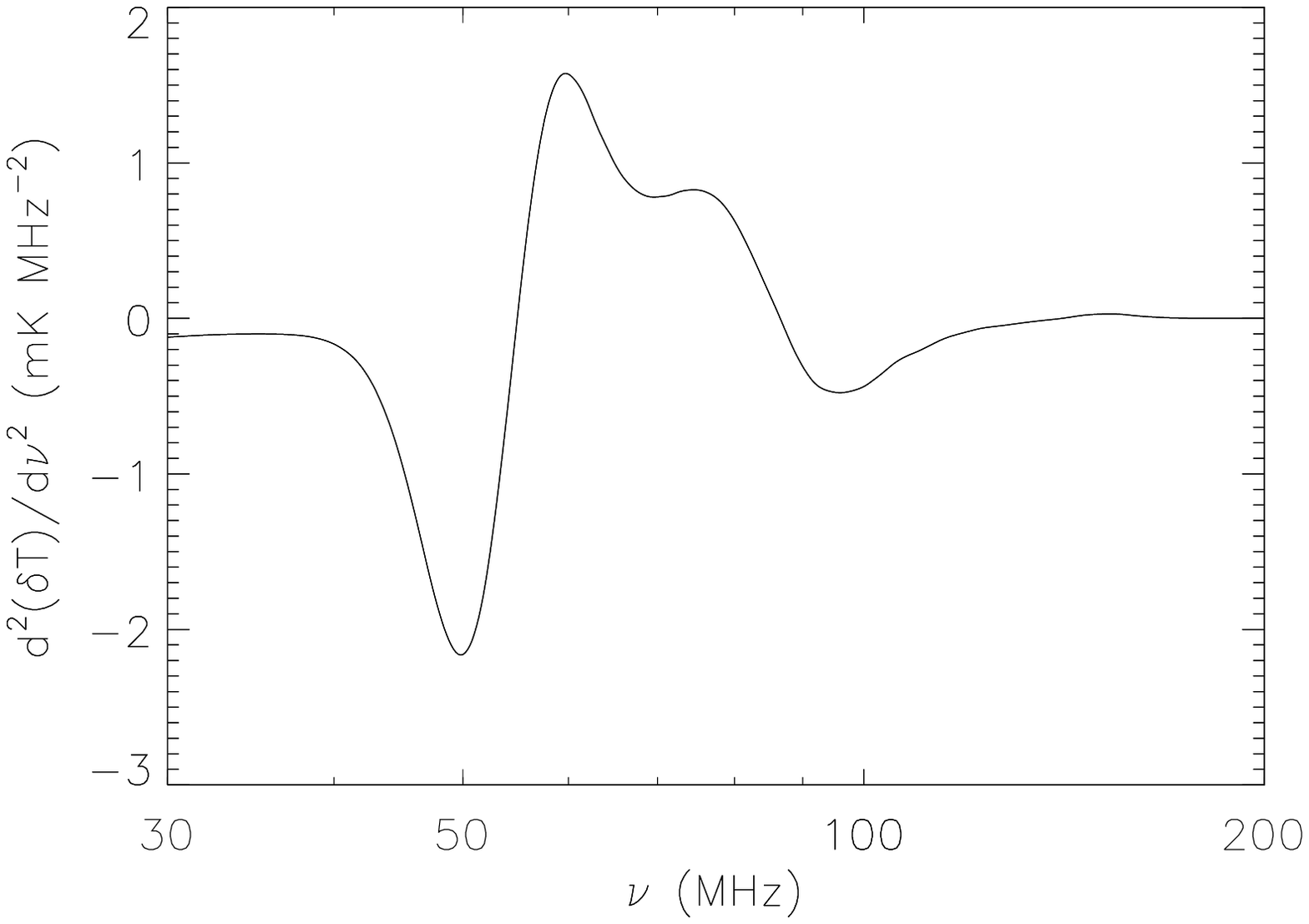,height=6.cm,width=9.cm,angle=0.0}
}
\end{center}
 \caption{\footnotesize{Results for the first model (solid line) of Fig.\ref{cmb_modified}.
 Upper panel: thermal SZE-21cm for $kT=7$ keV and $\tau=5\times10^{-3}$ calculated with the relativistic approach (solid line) and the non-relativistic approach (dashed line), compared with the standard SZE calculated with the relativistic approach (long-dashed line) and the non-relativistic approach (dotted line).
Middel panel: percentage difference between the relativistic result and the non-relativistic one.
Lower panel: second derivative of the input spectrum. 
 }}
 \label{app_fig1}
\end{figure}

\begin{figure}[ht]
\begin{center}
%\vspace{-10cm}
{
 \epsfig{file=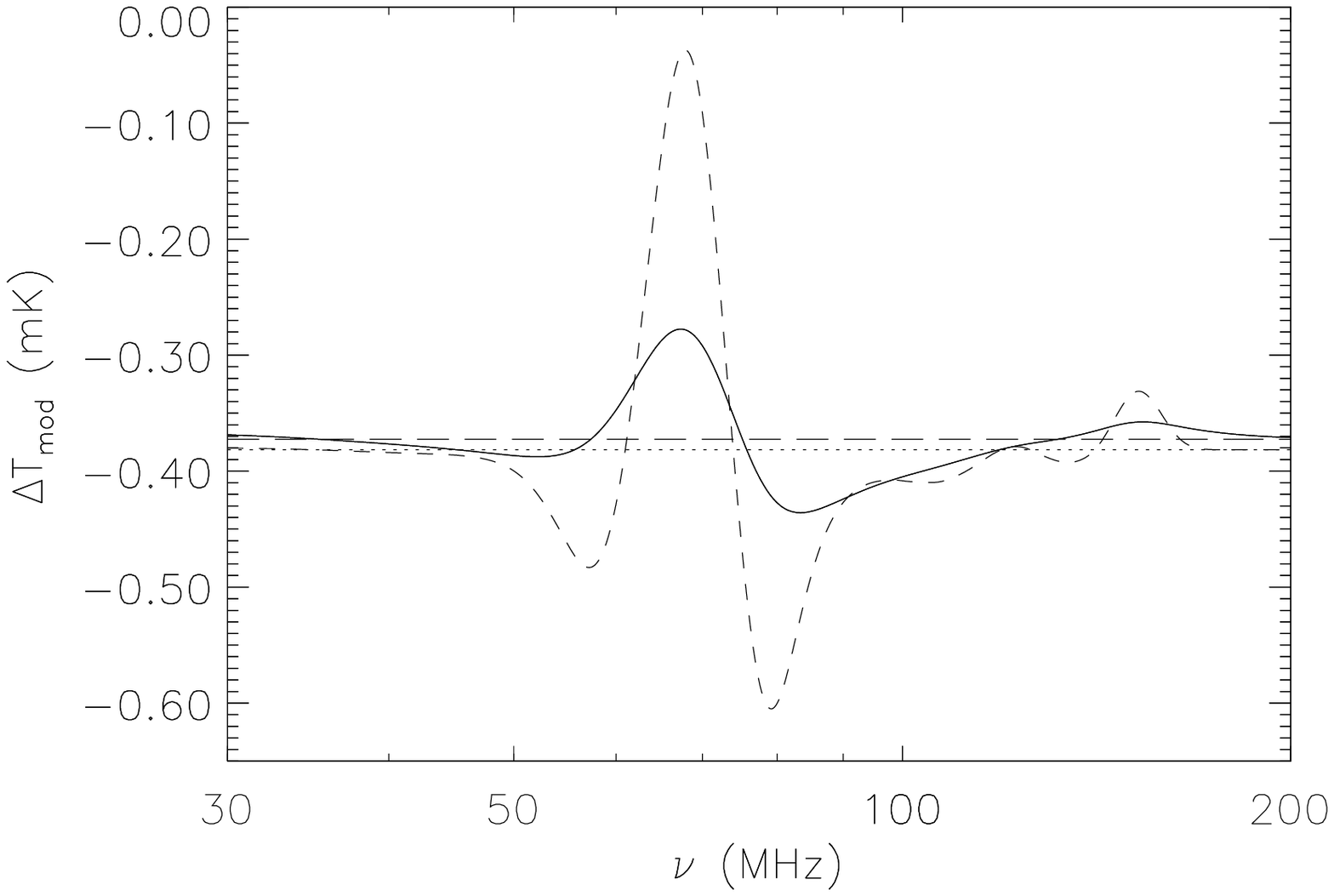,height=6.cm,width=9.cm,angle=0.0}
 \epsfig{file=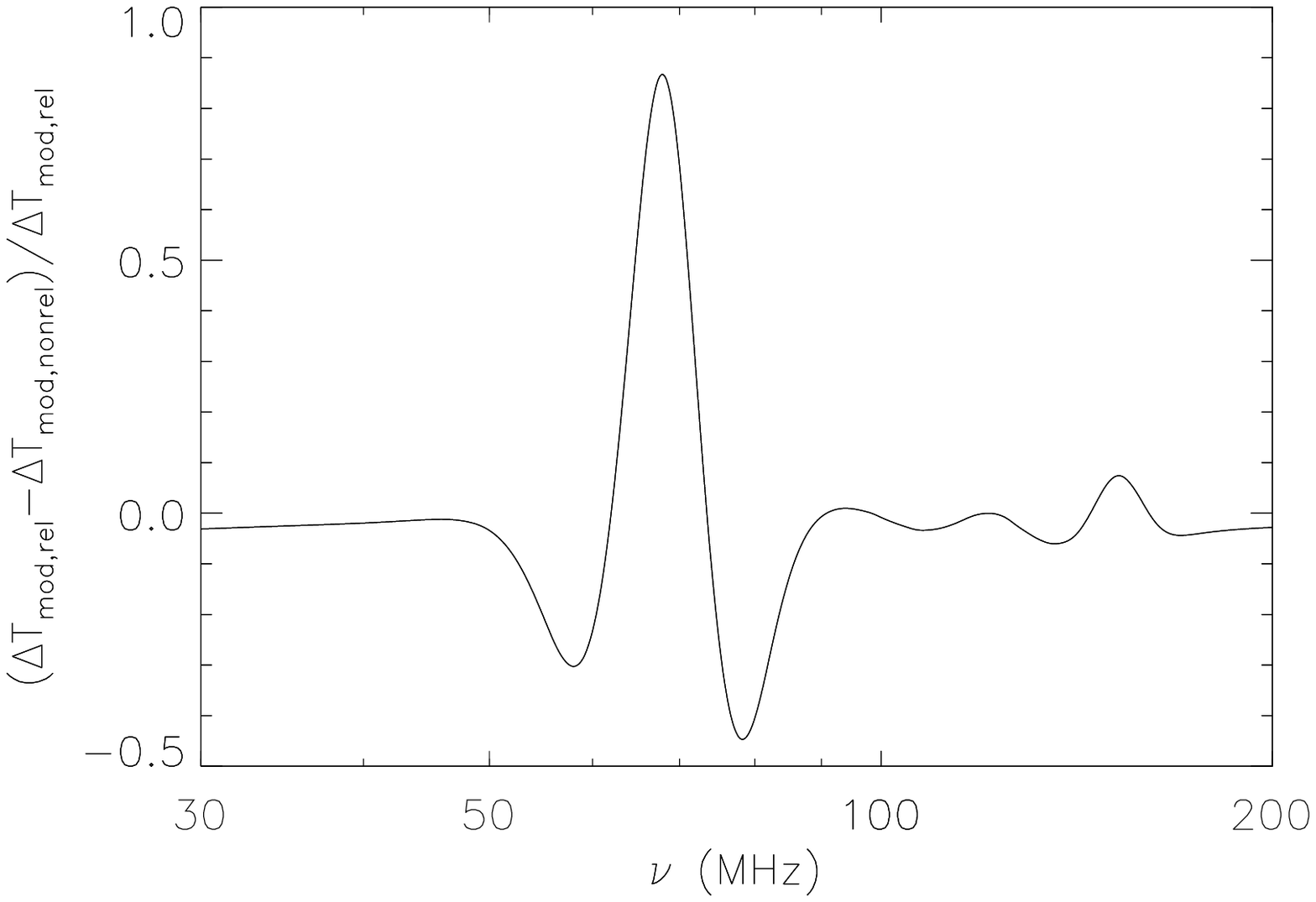,height=6.cm,width=9.cm,angle=0.0}
 \epsfig{file=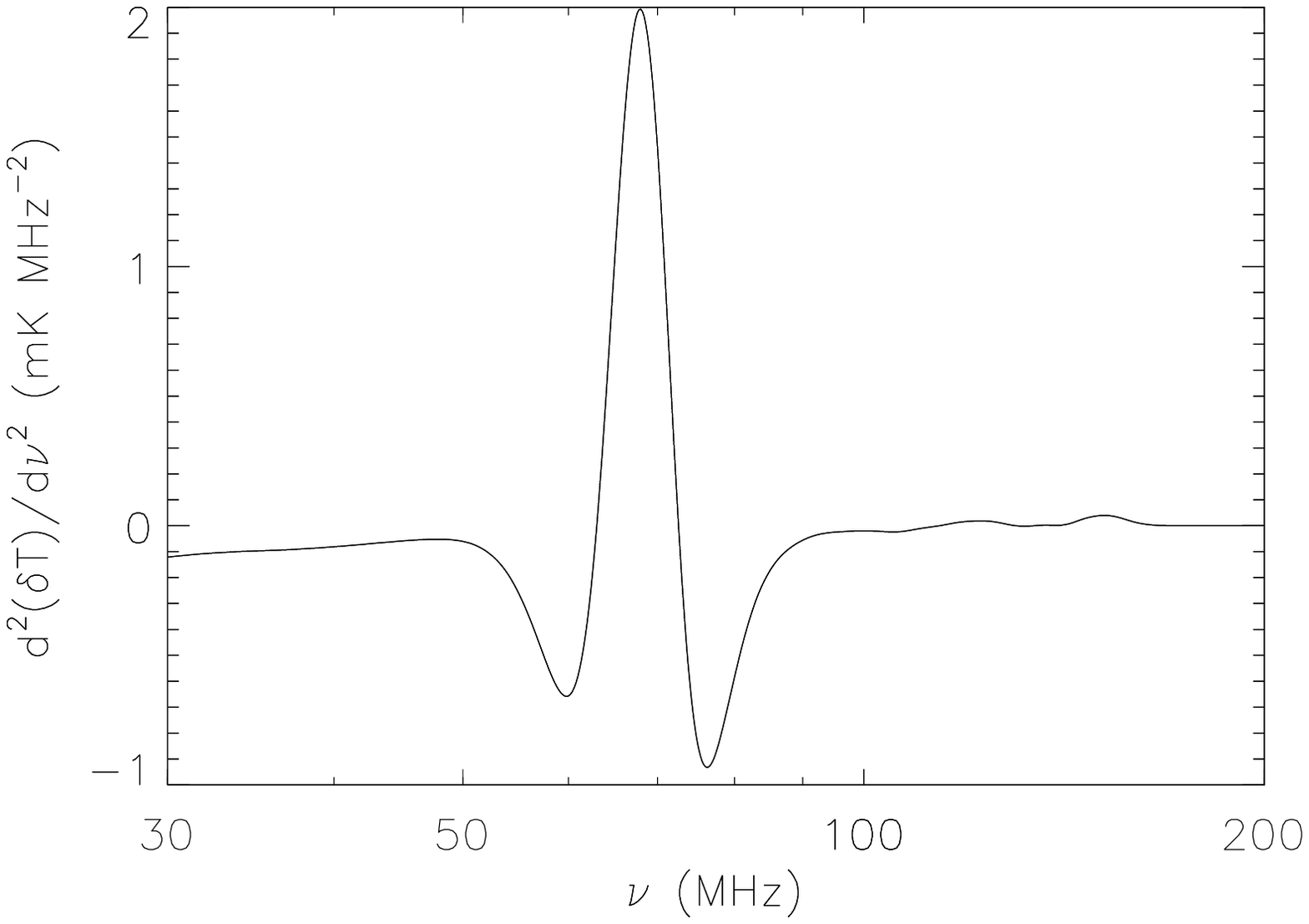,height=6.cm,width=9.cm,angle=0.0}
}
\end{center}
 \caption{\footnotesize{Like Fig.\ref{app_fig1} for the second model (dashed line) of Fig.\ref{cmb_modified}.
 }}
 \label{app_fig2}
\end{figure}

\begin{figure}[ht]
\begin{center}
%\vspace{-10cm}
{
 \epsfig{file=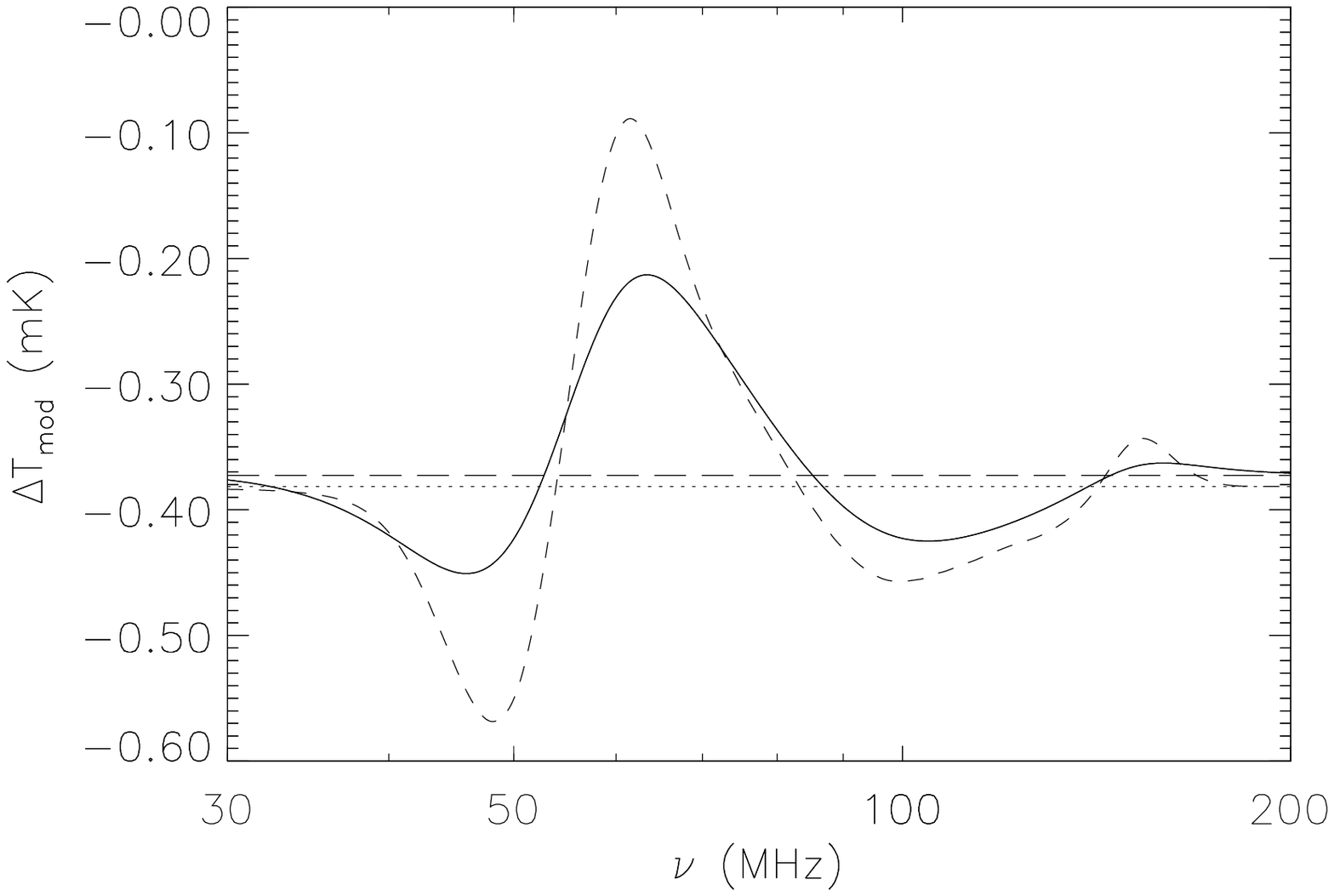,height=6.cm,width=9.cm,angle=0.0}
 \epsfig{file=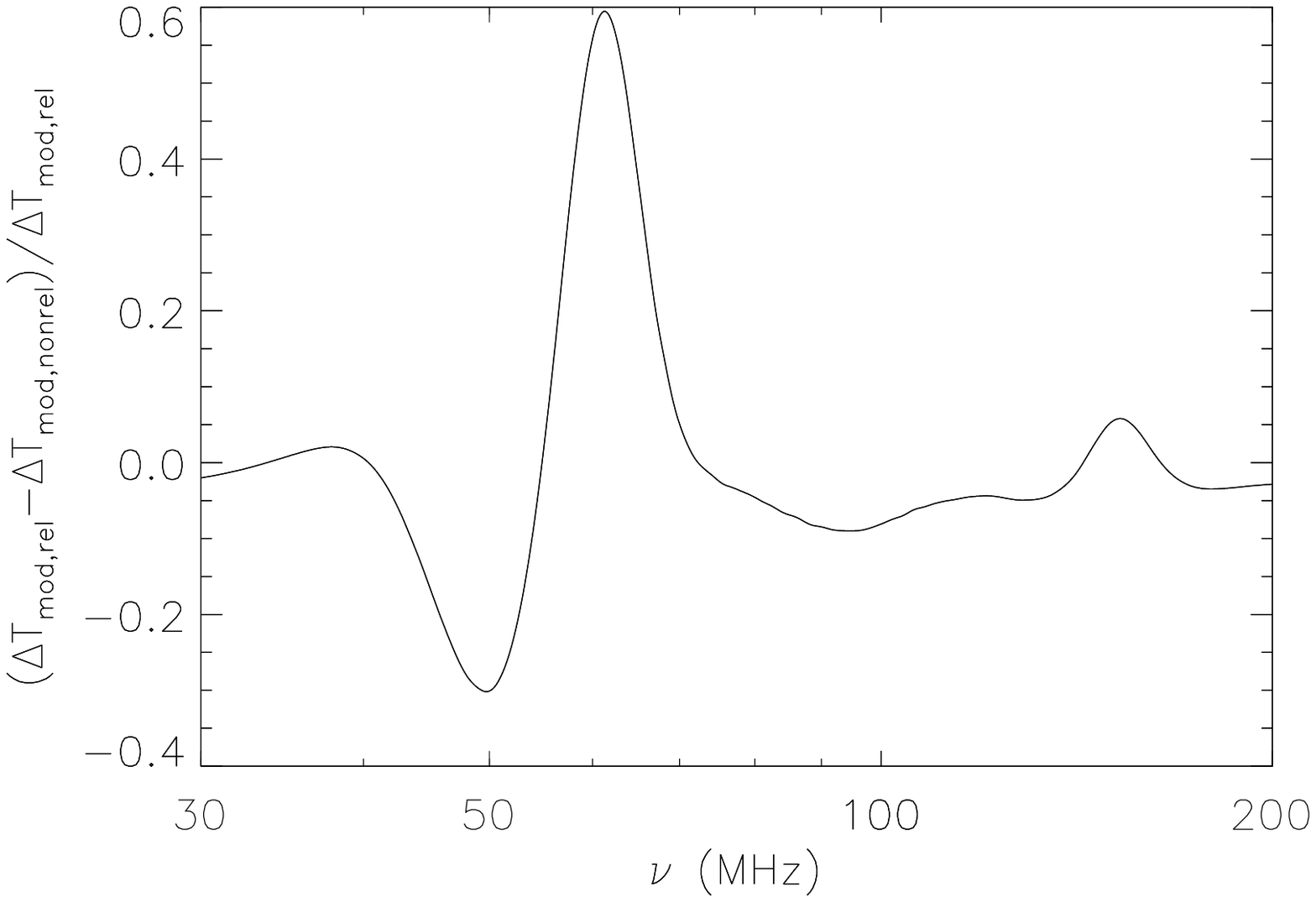,height=6.cm,width=9.cm,angle=0.0}
 \epsfig{file=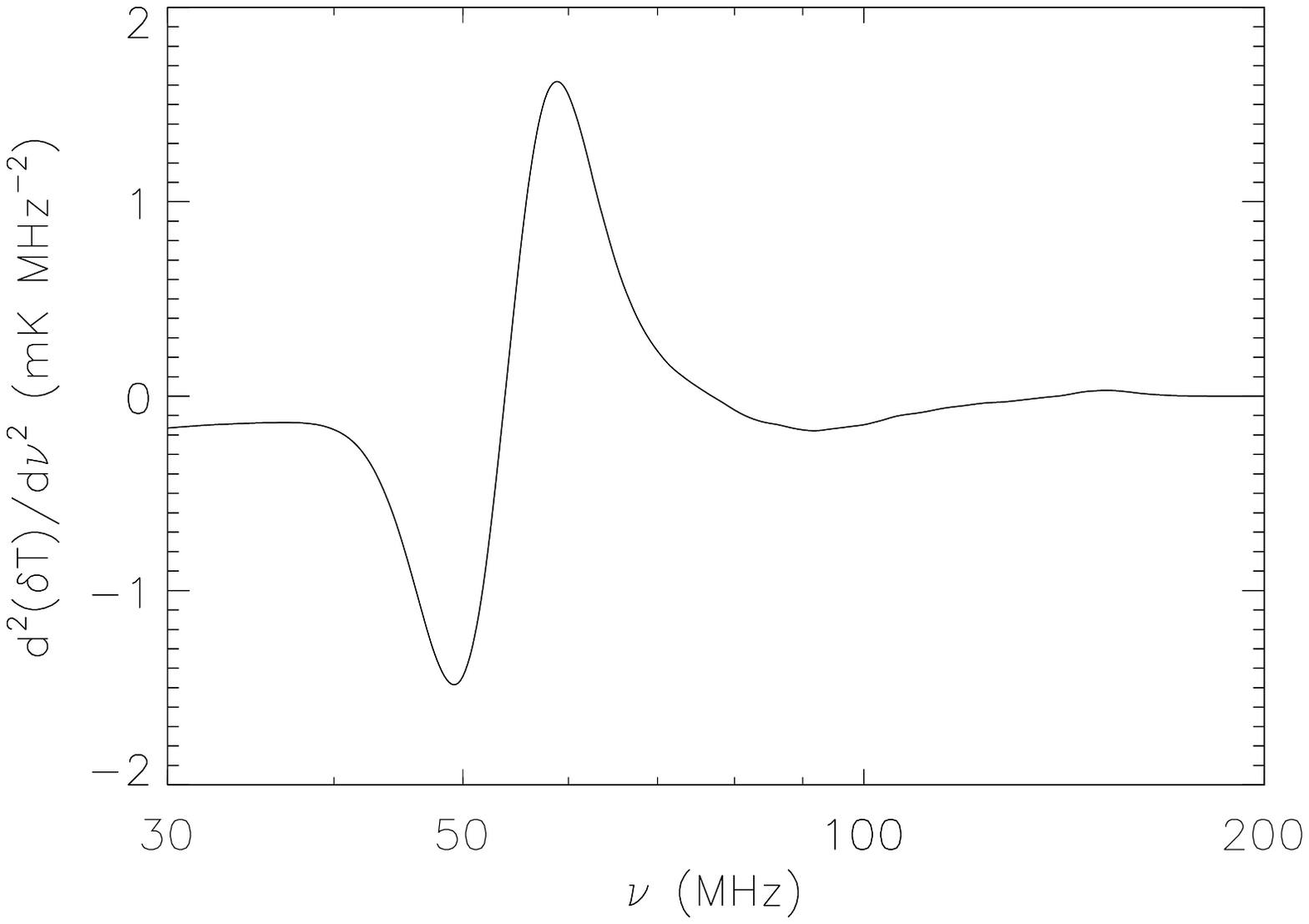,height=6.cm,width=9.cm,angle=0.0}
}
\end{center}
 \caption{\footnotesize{Like Fig.\ref{app_fig1} for the third model (dot-dashed line) of Fig.\ref{cmb_modified}.
 }}
 \label{app_fig3}
\end{figure}

\begin{figure}[ht]
\begin{center}
%\vspace{-10cm}
{
 \epsfig{file=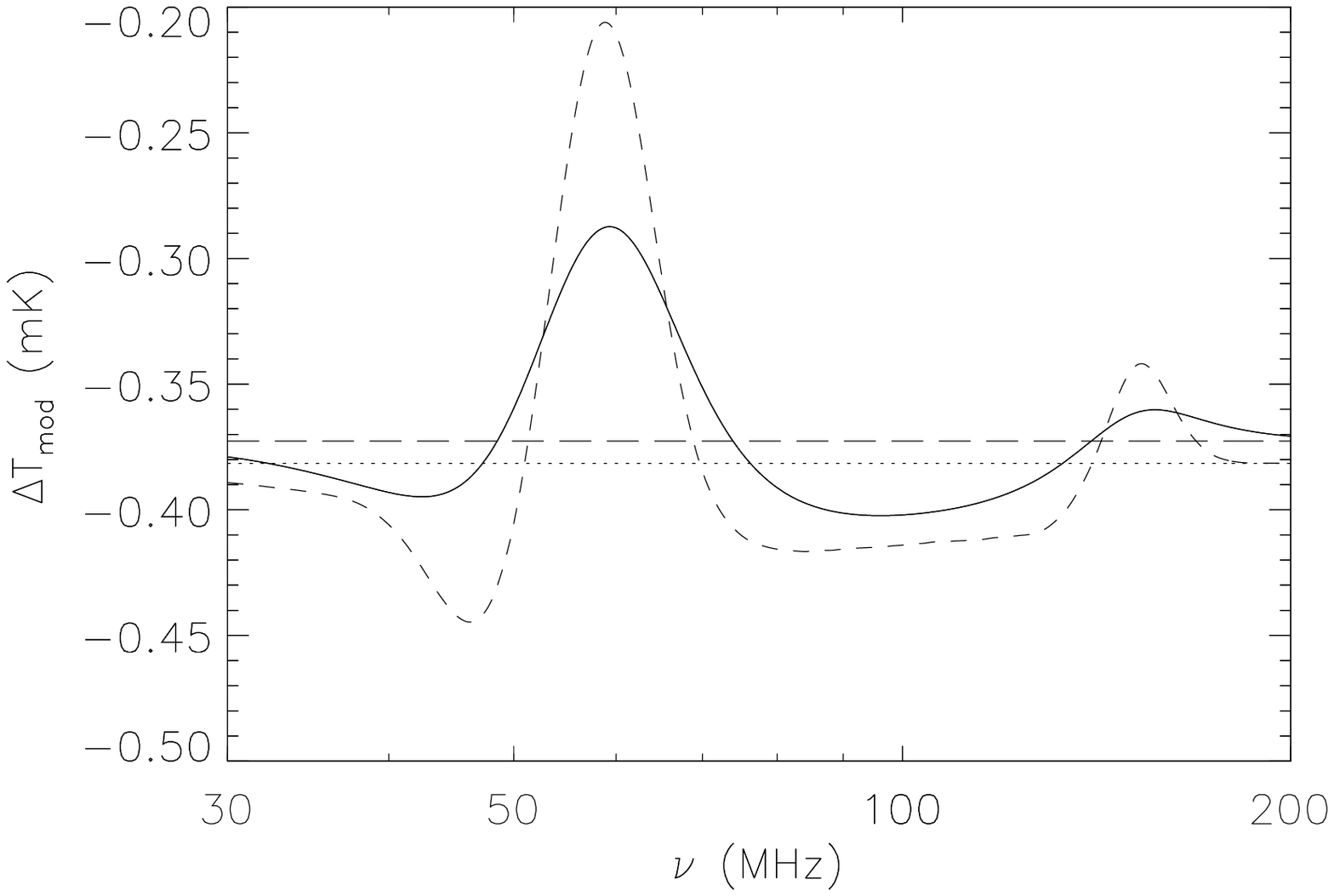,height=6.cm,width=9.cm,angle=0.0}
 \epsfig{file=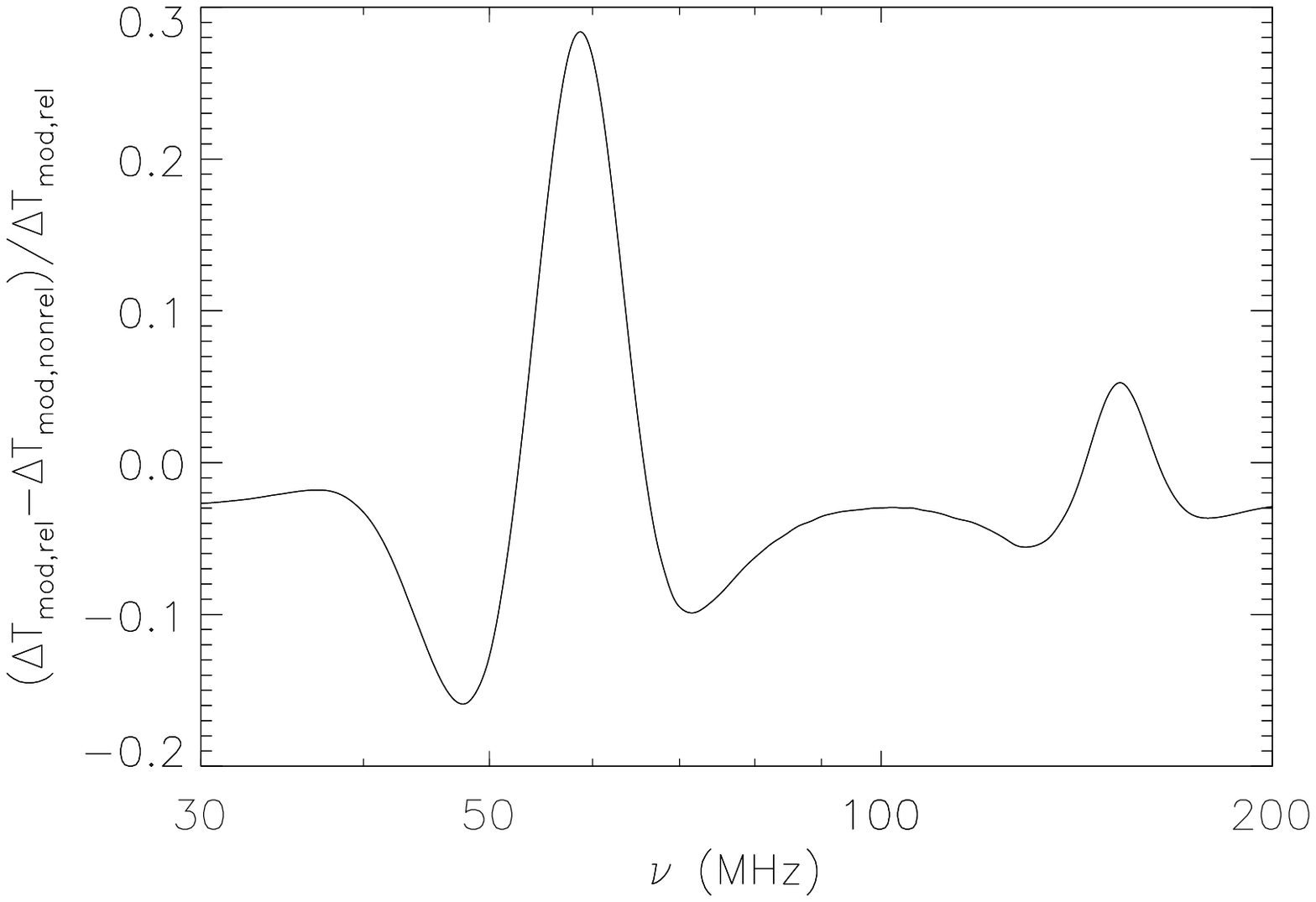,height=6.cm,width=9.cm,angle=0.0}
 \epsfig{file=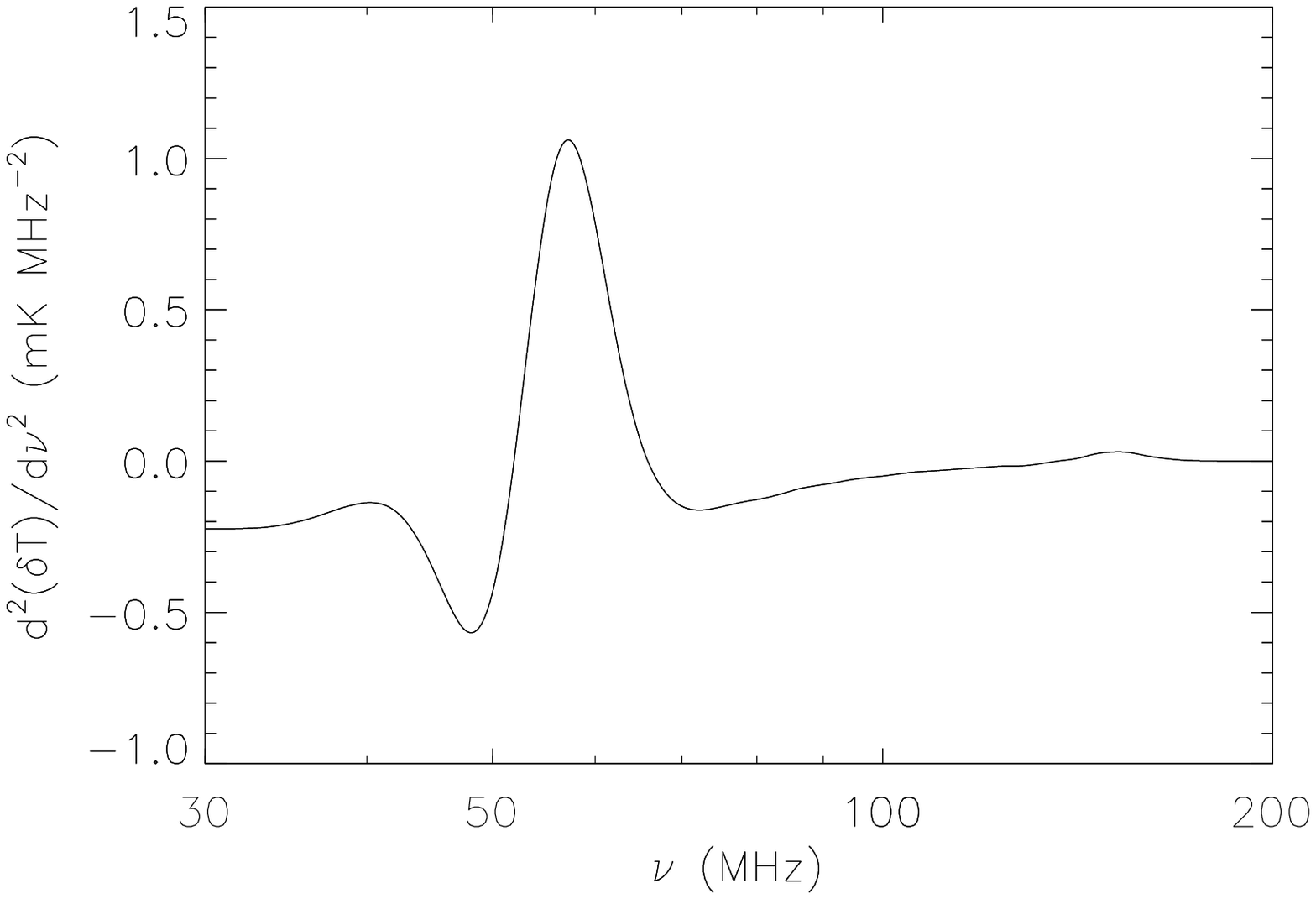,height=6.cm,width=9.cm,angle=0.0}
}
\end{center}
 \caption{\footnotesize{Like Fig.\ref{app_fig1} for the fourth model (three dots-dashed line) of Fig.\ref{cmb_modified}.
 }}
 \label{app_fig4}
\end{figure}

To discuss this issue, we show the spectra of the SZE-21cm calculated with the relativistic and the non-relativistic approach for the four input models we are using in this paper, and for a reference electron temperature of 7 keV (see Figs.\ref{app_fig1}-\ref{app_fig4}). The SZE-21cm is also compared with the standard SZE calculated with the relativistic and the non-relativistic approach. We show that the use of the non-relativistic approach introduces an overall numerical error into the standard SZE, and that this error is amplified in a frequency-dependent way for the SZE-21cm.

To better study the frequency dependence  of this error, we also show the percentage error done  in these cases, and we compare these results with the properties of the input spectrum. 
As discussed in Sect.3, we expect that the most important factor in determining the error done with the non-relativistic approach is the curvature of the input radiation spectrum: if the input spectrum has a large curvature this implies that using a function $P(s)$ that is narrower than the correct relativistic one (like in the non-relativistic approach) gives a result that is more different from the correct one w.r.t. the case where the input radiation spectrum is smooth, like in the case of the standard CMB. To check this conclusion, we compare the percentage error for the four models with the second derivative of the input radiation spectrum.

As expected, we observe that the percentage difference between the relativistic and non-relativistic spectrum has maximum points lying at frequencies where there is a peak  in the second derivative of the input radiation spectrum, corresponding to point of maximum curvature.\\ 
For the first model, we observe that there are two peaks in the frequency range 60--80 MHz in the case of the non-relativistic SZE-21cm. The existence of these peaks depends on the fact that the input radiation spectrum has two peaks in its second derivative, and the effect of using the non-relativistic Kernel introduces numerical artifacts due to the fact we are convolving the input radiation spectrum with a very narrow Kernel (see Birkinshaw et al. 1999). With the correct relativistic Kernel, the input spectrum is convolved with a wider function and the two peaks are then smoothed in only one peak. Therefore, the use of a non-relativistic approach gives origin not only to a numerical error in the value of the computed SZE, but also in its spectral shape and this error increases for increasing electron temperatures.

In the other models we consider in our paper, the second derivative of the input radiation spectrum has only one peak at frequencies $\nu\sim60-70$ MHz, and as a result also the non-relativistic SZE-21cm has only one peak in this spectral range. It can be seen that there are peaks/troughs in the percentage difference at frequencies whereby peaks/troughs are in the second derivative of the input spectrum (e.g. at $\nu\sim153$ MHz for the second model). This shows that the smoothness of the input radiation spectrum is an important aspect which produces differences in computing the SZE spectrum using a relativistic approach or a non-relativistic approach.

To conclude, we have shown in this Appendix that there is a substantial numerical error when computing the SZE using a non-relativistic approach, in particular when the input radiation spectrum is not a smooth function, as in the case of the modified CMB giving rise to the 21-cm background. This means that when using SZE of cosmic structures to study the cosmological 21-cm, it is imperative to use a full relativistic computation in order to obtain the correct SZE amplitude and its spectral shape.

\begin{thebibliography}{}

\bibitem{Barkana2001}
Barkana, R. \& Loeb, A., 2001, Phys. Rept., 349, 125

\bibitem{Barkana2005a}
Barkana, R. \& Loeb, A., 2005a, ApJ, 624, L65

\bibitem{Barkana2005b}
Barkana, R. \& Loeb, A., 2005b, ApJ, 626, 1

\bibitem{Bharadwaj2004}
Bharadwaj, S., \& Ali, S.S., 2004, MNRAS, 352, 142

\bibitem{Birkinshaw1999}
Birkinshaw, M. 1999, Phys. Rep., 310, 97

\bibitem{Bromm2004}
Bromm, V., \& Larson, R.B., 2004, ARA\&A, 42, 79

\bibitem{Carilli2004}
Carilli, C.L., et al., 2004, NewAR, 48, 1029

\bibitem{Cavaliere1976}
Cavaliere, A. \& Fusco-Femiano, R, 1976, A\&A, 49, 137

\bibitem{Chen2004}
Chen, X.L., \& Miralda-Escude, J., 2004, ApJ, 602, 1

\bibitem{Choudhury2006}
Choudhury, T.R., \& Ferrara, A., 2006, arXiv:astro-ph/0603149

\bibitem{Ciardi2005}
Ciardi, B., \& Ferrara, A., 2005, SSRv, 116, 625

\bibitem{Colafrancesco2013}
Colafrancesco, S. 2013, Acta Pol., 53, 560

\bibitem{Colafrancesco2014}
Colafrancesco, S. \& Marchegiani, P., 2014, A\&A, 562, L2

\bibitem{Colafrancescoetal2003}
Colafrancesco, S., Marchegiani, P. and Palladino, E., 2003, A\&A, 397, 27-52

\bibitem{Colafrancescoetal2011}
Colafrancesco, S., Marchegiani, P. \& Buonanno, R., 2011, A\&A, 527, L1

\bibitem{Colafrancescoetal2015}
Colafrancesco, S., Emritte, M.S., \& Marchegiani, P., 2015, JCAP, 05, 006

\bibitem{Cooray2004}
Cooray, A., 2004, Phys. Rev. D, 70, 063509

\bibitem{Cooray2006}
Cooray, A., 2006, Phys. Rev. D, 73, 103001

\bibitem{deOliveira2008}
de Oliveira-Costa, A. et al., 2008, MNRAS, 388, 247

\bibitem{ska2012}
Dewdney, P., Turner, W., Millenaar, R., McCool, R., Lazio, J. \& Cornwell, T., 2012, SKA baseline design document, {\slshape  http://www.skatelescope.org/wp-content/uploads/2012/07/SKA-TEL-SKO-DD-001-1\_BaselineDesign1.pdf}

\bibitem{Dillon2014}
Dillon, J.S. et al., 2014, Phys. Rev. D, 89, 023002

\bibitem{Ensslin2000}
En\ss lin, T.A. \& Kaiser, C.R., 2000, A\&A, 360, 417

\bibitem{Evoli2014}
Evoli, C., Mesinger, A., \& Ferrara, A., 2014, JCAP, 11, 024

\bibitem{Field1959}
Field, G.B.,  1959, ApJ, 129, 536

\bibitem{Furlanetto2004}
Furlanetto, S.R., \& Briggs, F.H., 2004, NewAR, 48, 1039

\bibitem{Furlanetto2006}
Furlanetto, S.R., Oh, S.P., \& Briggs, F.H., 2006, Phys. Rep., 433, 181

\bibitem{Liu2013}
Liu, A., et al., 2013, Phys. Rev. D, 87, 043002

\bibitem{Loeb2001}
Loeb, A., \& Barkana, R., 2001, ARA\&A, 39, 19

\bibitem{LoebZal2004}
Loeb, A., \& Zaldarriaga, M., 2004, Phys. Rev. Lett. 92, 211301

\bibitem{McQuinn2006}
McQuinn, M., et al., 2006, ApJ, 653, 815

\bibitem{Mesinger2011}
Mesinger, A., Furlanetto, S., \& Cen, R., 2011, MNRAS, 411, 955

\bibitem{Morales2010}
Morales, M.F., \& Wyithe, J.S.B., 2010, ARA\&A, 48, 127 

\bibitem{Paciga2013}
Paciga, G. et al., 2013, MNRAS, 433, 639

\bibitem{Parsons2013}
Parsons, A.R. et al. 2014, ApJ, 788, 106

\bibitem{Pritchard2010}
Pritchard, J.R., \& Loeb, A. 2010, Phys. Rev. D, 82, 023006

\bibitem{Pritchard2012}
Pritchard, J.R., \& Loeb, A. 2012, Rep. Progr. Phys., 75, 086901

\bibitem{Shaver1999}
Shaver, P.A.,  Windhorst, R.A., Madau, P. \& de Bruyn, A.G., 1999, A\&A, 345, 380 

\bibitem{ScottRees1990}
Scott, D. \& Rees, M.J. 990, MNRAS, 247, 510

\bibitem{Valdes2013}
Valdes, M., et al., 2013, MNRAS, 429, 1705

\bibitem{Wout1952}
Wouthuysen, S.A., 1952, AJ, 57, 31

\bibitem{Zaroubi2013}
Zaroubi, S.  2013, The First Galaxies, Astrophysics and Space Science Library, 396, 45


\end{thebibliography}
\end{document}